\documentclass[
11pt, 
english, 
singlespacing, 
headsepline, 
]{MastersDoctoralThesis} 

\usepackage[utf8]{inputenc} 
\usepackage[T1]{fontenc} 

\usepackage{palatino} 
\usepackage{amsmath}
\usepackage{amssymb}
\usepackage[backend=bibtex,style=phys,articletitle=false,biblabel=brackets,chaptertitle=false,pageranges=false, sorting=none]{biblatex} 

\usepackage{bm}
\usepackage[autostyle=true]{csquotes} 
\usepackage{lipsum}
\usepackage{comment}

\makeatletter
\renewenvironment{abstract}
{%
  \typeout{Abstract Page}%
  \cleardoublepage 
  \thispagestyle{empty}%
  \begingroup
    \centering
    \setlength{\parskip}{0pt}%
    {\normalsize \par}%
    \nobreak\bigskip
    {\underline{ABSTRACT} \par}%
    \nobreak\bigskip
    {\normalsize {}\par}%
    {\normalsize {} \par}%
    \nobreak\bigskip
    {\normalsize \par}%
    \nobreak\bigskip
    {\normalsize\bfseries  \par}%
    \nobreak\medskip
    {\normalsize \par}%
  \endgroup
  \nobreak\addvspace{\glueexpr\topsep+\partopsep}%
}{}

\makeatother
\def\ra{\rangle}

\newcommand{\ii}{\text{i}}

\thesistitle{Spontaneous Supersymmetry Breaking and Nambu-Goldstone Modes in Interacting Majorana Chains} 
\supervisor{Hosho \textsc{Katsura}} 
\examiner{} 
\degree{Ph. D of Science} 
\author{Noriaki \textsc{Sannomiya}} 
\addresses{} 

\subject{Physics} 
\university{\href{http://www.u-tokyo.ac.jp/}{University of Tokyo}} 
\department{\href{http://www.phys.s.u-tokyo.ac.jp/}{Department of Physics}} 
\group{\href{http://park.itc.u-tokyo.ac.jp/hkatsura-lab/}{Katsura Group}} 

\AtBeginDocument{
\hypersetup{pdftitle=\ttitle} 
\hypersetup{pdfauthor=\authorname} 
\hypersetup{pdfkeywords=\keywordnames} 
}

\begin{document}

\frontmatter 

\pagestyle{plain} 


\begin{titlepage}
\begin{center}

\vspace*{.06\textheight}
{\scshape\LARGE \univname\par}\vspace{1.5cm} 
\textsc{\Large Ph. D Thesis}\\[0.5cm] 

\HRule \\[0.4cm] 
{\huge \bfseries \ttitle\par}\vspace{0.4cm} 
\HRule \\[1.5cm] 
 
\begin{minipage}[t]{0.4\textwidth}
\begin{flushleft} \large
\emph{Author:}\\
\href{http://www.johnsmith.com}{\authorname} 
\end{flushleft}
\end{minipage}
\begin{minipage}[t]{0.4\textwidth}
\begin{flushright} \large
\emph{Supervisor:} \\
\href{http://www.jamessmith.com}{\supname} 
\end{flushright}
\end{minipage}\\[3cm]
 
\vfill

\large \textit{A thesis \\ for the degree of \degreename}\\[0.3cm] 
\textit{in the}\\[0.4cm]
\groupname\\\deptname\\[2cm] 
 
\vfill

{\large \today}\\[4cm] 

\vfill
\end{center}
\end{titlepage}


\begin{abstract}
\addchaptertocentry{\abstractname} 
In modern physics, spontaneous symmetry breaking is one of the most important and ubiquitous ideas. It is well known that the breakdown of global and continuous symmetry gives rise to gapless excitations called Nambu-Goldstone (NG) bosons. This is called NG theorem. Since the generators of broken symmetries are bosonic, the excitations are also bosons. On the other hand, excitations are thought of as fermionic when a generator of broken symmetry is fermionic. The most famous example of fermionic symmetry is supersymmetry (SUSY). SUSY is a symmetry that exchanges bosons and fermions. In high-energy physics, it is known that spontaneous SUSY breaking gives rise to massless modes called NG fermions or Nambu-Goldstinos. In non-relativistic systems, fewer examples of models with SUSY are studied. Therefore, spontaneous SUSY breaking in non-relativistic systems is less understood than that in nonrelativistic systems or spontaneous symmetry breaking. In this thesis, we introduce interacting Majorana models with supersymmetry (SUSY) and discuss the properties in terms of spontaneous SUSY breaking and NG modes. 

Chapter~\ref{chap:intro} gives an introduction to supersymmetry, $\mathcal N = 2$ supersymmetric lattice models, and the properties of Majorana fermions. In addition, we discuss the Kitaev chain in terms of the ground state. In Chapter~\ref{chap:N1susy}, we briefly review $\mathcal N = 1$ SUSY quantum mechanics and the definition of spontaneous SUSY breaking. 

In Chapter~\ref{chap:MajoNicolai}, we introduce and study an interacting Majorana fermion model with SUSY. This model has one parameter $g$, and $g$ connect strongly interacting limit and free Majorana fermion limit. Depending on the values of $g$, we study the properties of SUSY in terms of spontaneous SUSY breaking. When $g = 1$, we prove that SUSY is unbroken and the ground states are solvable. We find that SUSY is restored in the infinite volume limit in an extended area of the parameter $g$. When $g > 8/\pi - 1$, we prove that SUSY is broken spontaneously by giving a lower bound of the ground state energy density. When SUSY is spontaneously broken, we prove that there exist gapless modes in this model. Using exact diagonalization, we calculate the dispersion relation of Nambu-Goldstone modes and find that it is cubic in momentum.

In Chapter~\ref{chap:LNicolai}, we introduce and study another interacting Majorana fermion model with SUSY. This model also has one parameter $g$, and $g$ connect strongly interacting limit and free Majorana fermion limit. Depending on the values of $g$, we study the properties of SUSY in terms of spontaneous SUSY breaking, In the case of $g=1$, we prove that SUSY is unbroken and the ground states are integrable. We find that SUSY is restored in the infinite volume limit for modest values of $g$. When $g > 8/\pi - 1$, we prove that SUSY is spontaneously broken. We prove that there exist gapless modes associated with spontaneous SUSY breaking. Using exact diagonalization, we find that the dispersion relation of Nambu-Goldstone modes is linear in momentum.

In chapter~\ref{sec:conclusion}, we give a summary of this thesis. 
\end{abstract}


\begin{acknowledgements}
\addchaptertocentry{\acknowledgementname} 
{I would like to thank my supervisor, Hosho Katsura, for all his
help and support during my doctoral course. I can say that I could not wish for a better supervisor than him, and have learned a lot from him how to approach problems. I have huge admiration for his professionalism as a scientist.

I also would like to thank colleagues in the Katsura group who have provided a fun and interesting time. In particular, I'm very encouraged by Akagi-san, an assistant professor of the Katsura group, and Kondo-san, Shibata-san, and Yang-san, graduate students of the Katsura group.

Finally, I would like to thank my parents and grandmother for supporting me and always standing by me, through both my doctoral course and everything else.}

\end{acknowledgements}


\tableofcontents 

\mainmatter 

\pagestyle{thesis} 



\chapter{Introduction} 

\label{chap:intro} 

In this chapter, we first provide a short history of supersymmetry (SUSY) and the birth of supersymmetric quantum mechanics (non-relativistic SUSY). Following this, we see general properties of non-relativistic SUSY and special properties of the Nicolai model which is a model with non-relativistic SUSY. 

\section{Supersymmetry}
Symmetry plays an important role in many areas of physics. In nature, symmetries are sometimes broken spontaneously. Spontaneous symmetry breaking (SSB) is defined naively as $Q|{\rm gs}\rangle \neq 0$, where $Q$ and $|{\rm gs}\rangle$ are the generator of a symmetry and the ground state of the system, respectively. For the latter purpose, we would like to give a more precise definition of spontaneous symmetry breaking. If there exists an operator $\psi(x)$ which satisfies the following relation
\begin{align}
\langle\phi|[Q,\psi(x)]|\phi\rangle\neq0, \label{eq:defSSB}
\end{align}
the symmetry is said to be spontaneously broken. Here, $|\phi\rangle$ is the ground state,  $Q$ is a generator of symmetry.
We note that SSB occurs only in the infinite-volume limit. Famous examples of such SSB are translational symmetry breaking in solids and spin rotational symmetry breaking in magnets. It is known that there exist gapless modes which satisfy $E_{\bm k}\to 0$ as ${\bm k}\to 0$ known as Nambu-Goldstone (NG)  bosons when a global and continuous symmetry is spontaneously broken \cite{PR_Nambu,NC_Goldstone, PR_Goldstone}. This is called Nambu-Goldstone theorem. This theorem is applicable to both relativistic and non-relativistic systems. When the system holds Lorentz symmetry, the number of NG bosons is the same as the number of broken symmetries. In non-relativistic systems, classification of NG modes in terms of the ground state expectation values of commutation relation of generators of symmetries determines the number of NG modes \cite{PRL_Watanabe, PRL_Hidaka}.

The above SSB is mainly discussed in the context of bosonic symmetry. On the other hand, supersymmetry (SUSY) is a symmetry that relates bosonic particles and fermionic particles. SUSY is considered to solve the hierarchy problem that is an important topic in high energy physics. However, despite a great deal of effort, SUSY has yet to be experimentally discovered. In this section, we briefly review the history of SUSY.

This symmetry treats bosons and fermions as a multiplet. The algebra of this symmetry is not a Lie algebra since it involves both commutator and anti-commutator. This symmetry does not contain the  Lorentz group. In 1967, Coleman and Mandula showed a theorem which states that a possible symmetry commuting with the S-matrix is restricted to translation, Lorentz rotation, or internal symmetries~\cite{PR_Coleman}.  SUSY is not contained in these three symmetries since SUSY changes integer spin particles into half-integer ones while the three symmetries do not. In 1974, Wess and Zumino constructed a (3+1) dimensional model which has both SUSY and Lorentz symmetry~\cite{LPB_Wess}. Coleman-Mandula's no-go theorem is applicable only to bosonic generators of symmetries. On the other hand, the theorem cannot be applicable to fermionic generators of symmetries such as those of SUSY. In 1975, Haag, \L opusza\'nski and Sohnius generalized Coleman-Mandula's no-go theorem to that containing SUSY,  and they showed that generators of SUSY must satisfy superalgebra~\cite{NPB_Haag}, 
\begin{align}
\{Q_a,\bar{Q}_b\} & =2(\gamma^\mu)_{ab}P_\mu \ , \\
[Q_a,P_\mu] & =0 \ ,\\
[Q_a,M_{\mu\nu}] & =(\sigma^4_{\mu\nu})_{ab}Q_b,
\end{align}
where $Q_a$, $P_\mu$, and $M_{\mu\nu}$ are generators of SUSY, translational symmetry, and Lorenz rotational symmetry,  respectively. The index $a$ ($b$) stands for the spinor ($a,b=1,...,4$), and $\mu$ ($\nu$) is index of space time ($\mu,\nu=0,...,3$). The operator $\sigma^4_{\mu\nu}$ is defined as a commutator of gamma matrices, $\sigma^4_{\mu\nu}=(\mathrm{i}/4)[\gamma_\mu,\gamma_\nu]$. The operator $\bar{Q}_a$ is defined as $\bar{Q}_a=(Q^\dagger \gamma_0)_a$. Other generators of symmetries, including $P_\mu$ and $M_{\mu\nu}$, satisfy the same algebra as non-supersymmetric theories. We comment in passing on the work by Akulov and Volkov. They constructed a theory which has SUSY before the work by Wess and Zumino. This is a non-linear realization of SUSY and describes Goldstone fermions brought about by SUSY breaking~\cite{PZETF_Volkov,PLB_Volkov}. The paper \cite{PZETF_Volkov} was written in Russian so that the mainstream of the early development of SUSY was independent of the work by Akulov and Volkov.

Nicolai introduced a model of supersymmetric quantum mechanics (SUSY QM) in 1976~\cite{JPA_Nicolai_76}. This is the first model of SUSY QM~\cite{JPA_Junker}. The model describes the interacting spin chain with SUSY. Five years later, Witten introduced the SUSY QM model as a toy model of dynamical SUSY breaking in quantum field theory~\cite{Witten_NPB81}. Then he introduced a topological index which is known as the "Witten index" in a subsequent paper~\cite{Witten_NPB82}. A model introduced by Witten treats one particle physics in a continuous system.

Next, we comment on spontaneous SUSY breaking in relativistic systems. SUSY breaking is naively defined as~\cite{Kirsten}
\begin{align}
Q_A|{\rm vac}\rangle\neq0, \quad \bar{Q}_{\dot A}|{\rm vac}\rangle\neq0.
\label{eq:Qvac}
\end{align}
Here, $Q_A$ is the supercharge, the operator $\bar{Q}_{\dot A}$ is defined by $\bar{Q}_{\dot A}=(Q_A)^\dagger$, $A$ is the index referring to spinor ($A=1,2$), and $|{\rm vac}\ra$ is a vacuum of the system.
From superalgebra, the Hamiltonian becomes the following form~\cite{Kirsten},
\begin{align}
H=\frac{1}{4}\left\{Q_1\bar{Q}_{\dot{1}}+\bar{Q}_{\dot{1}}Q_1+Q_2\bar{Q}_{\dot{2}}+\bar{Q}_{\dot{2}}Q_2\right\}.
\end{align}
One can find that the expectation value of the energy in the vacuum becomes
\begin{align}
\langle{\rm vac}|H|{\rm vac}\rangle\propto\sum_{A=1,2}|Q_A|{\rm vac}\rangle|^2+|\bar{Q}_{\dot{A}}|{\rm vac}\rangle|^2.
\label{eq:Hvac}
\end{align}
From Eqs. (\ref{eq:Qvac}) and (\ref{eq:Hvac}), we see that SUSY breaking is equivalent to the positivity of the vacuum energy.

Spontaneous SUSY breaking also gives rise to massless excitations called Goldstinos or Nambu-Goldstone fermions. Using the analogy of the Goldstone theorem, NG fermions are discussed in~\cite{PLB_Salam}. It is known that SUSY breaking occurs in both finite and the infinite systems while usual spontaneous symmetry breaking occurs only in the infinite system~\cite{Witten_NPB82}.


\section{${\mathcal N}=2$ Supersymmetric lattice model in non-relativistic systems}
\label{sec:SUSYQM}

Let us consider some general properties of ${\mathcal N}=2$ SUSY QM. Suppose there exist two supercharges $Q$ and $Q^\dagger$, which are fermionic operators and are nilpotent,
\begin{align}
Q^2=(Q^\dagger)^2=0. 
\label{eq:nilpotent}
\end{align}
The Hamiltonian is defined by the anti-commutator of the supercharges,
\begin{align}
H=\{Q,Q^\dagger\}. 
\label{eq:suham}
\end{align}
From the nilpotency of the supercharges, one can see that the supercharges are conserved,
\begin{align}
[H,Q]=[H,Q^\dagger]=0.
\label{eq:SuQcon}
\end{align}

By the definition of the Hamiltonian, all eigenvalues of the Hamiltonian are non-negative,
\begin{align}
\langle\psi|H|\psi\rangle=\parallel Q^\dagger|\psi\rangle\parallel^2+\parallel Q|\psi\rangle\parallel^2\ge0.
\label{eq:positive}
\end{align}
When the eigenvalue of the Hamiltonian is positive, all states come in pair, ($|\psi\rangle$, $Q^\dagger|\psi\rangle$) such that $Q|\psi\rangle=0$. The state $Q^\dagger|\psi\rangle$ is called a superpartner of $|\psi\rangle$, and vice versa. We can always choose $|\psi\ra$ to be annihilated by $Q$~\cite{PRL_Fendley_2003}. This can be seen as follows. Let $|\psi_0\ra$ be an eigenstate of $H$ with energy $E>0$ and suppose $Q|\psi_0\ra \ne 0$. Since $H$ commutes with $Q$ and $Q^\dagger$, 
\begin{equation}
|\psi\ra := |\psi_0\ra - \frac{1}{E}Q^\dagger Q|\psi_0\ra,
\end{equation}
is also an eigenstate of $H$ with the same energy. It then follows from $Q^2=0$ that $Q|\psi\ra=0$. 
All states with zero energy are singlet, and are annihilated by both $Q$ and $Q^\dagger$,
\begin{align}
Q|0\rangle=Q^\dagger|0\rangle=0.
\end{align}
Here, $|0\rangle$ is a state with zero energy. 
In addition to the conservation of the supercharges and nilpotency of them, models with SUSY satisfy the following algebra,
\begin{align}
\{(-1)^F,Q\}=\{(-1)^F,Q^\dagger\}=0.
\label{eq:pari}
\end{align}
Here, the operator $F$ is the total fermion number operator. These three algebras Eqs. (\ref{eq:nilpotent}, \ref{eq:SuQcon}, \ref{eq:pari}) is called superalgebra. In non-relativistic theory, the Poincar\'e invariance is violated. Therefore, the superalgebra of non-relativistic theory is simpler than that of relativistic theory.
From the relation Eq. (\ref{eq:pari}), we find that the SUSY QM model has a manifest $\mathbb{Z}_2$ symmetry, 
\begin{align}
[H,(-1)^F]=0.  
\end{align}
Equation (\ref{eq:pari}) also tells us that any positive-energy eigenstate $|\psi\ra$ and its superpartner $Q^\dagger|\psi\rangle$ have opposite fermionic parity,
\begin{align}
(-1)^F|\psi\ra=\pm|\psi\ra \Rightarrow (-1)^F(Q^\dagger|\psi\rangle)=\mp(Q^\dagger|\psi\rangle).
\end{align}
In this sense, the supercharge $Q^\dagger$ maps bosonic (fermionic) states into fermionic (bosonic) states. The operator $Q$ plays the same role. In condensed matter physics, $\mathcal{N}=2$ SUSY is discussed in the context of integrable systems \cite{PRL_Fendley_2003,PRL_Fendley_2005, EPJB_huijse, PRL_Huijse, NJP_Huijse} and cold atomic systems \cite{PRA_Yang, PRL_Yu, PRA_Shi, PRA_Bradlyn, PRA_Hidaka, PRA_Lai, PRA_Blaizot_2017, PRR_Tajima}. 

\begin{figure}[htb]
\begin{center}
\includegraphics[width=1.0\columnwidth]{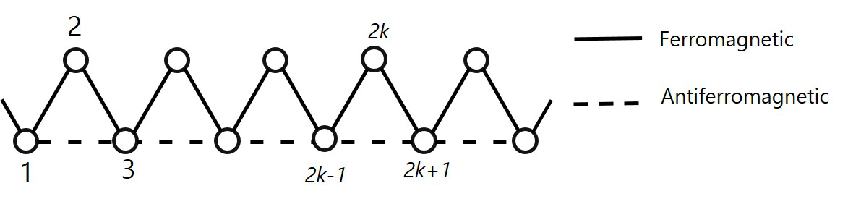}
\end{center}
\caption{A sketch of $\Delta$ chain. Each circle denotes a site.  The solid and dashed bonds represent ferromagnetic and antiferromagnetic
Ising interactions, respectively.}
\label{fig:deltachain}
\end{figure}

To review the spontaneous SUSY breaking and NG modes in ${\mathcal N} = 2$ SUSY, we introduce the Nicolai model~\cite{JPA_Nicolai_76, JPA_Nicolai_77}. This model was introduced to realize SUSY in the Hamiltonian formalism. The supercharge of the Nicolai model is given as
\begin{align}
Q=\sum_{l=1}^{N/2}c_{2l-1}c_{2l}^\dagger c_{2l+1},
\end{align}
where $c_j$ and $c_j^\dagger$ are annihilation and creation operators of spinless fermions on $j$-th site. The Hamiltonian is defined as $H=\{Q,Q^\dagger\}$, and its explicit form is given by
\begin{align}
H = & \sum^{N/2}_{k=1} (n_{2k} + n_{2k-1} n_{2k+1}) - \sum^N_{j=1} n_j n_{j+1} + \sum_{k=1}^{N/2} 
(c^\dagger_{2k}c^\dagger_{2k+3}c_{2k-1}c_{2k+2} +{\rm H.c.}).
\end{align}
Here, the operator $n_j$ is a fermion number operator at the $j$-th site and defined by $n_j:=c_j^\dagger c_j$.
For later purposes, we also provide a spin-model representation of $H$,
\begin{align}
H= & \frac{1}{4}\sum_{k=1}^{N/2}(1+\sigma_{2k-1}^z\sigma_{2k+1}^z-\sigma_{2k-1}^z\sigma_{2k}^z-\sigma_{2k}^z\sigma_{2k+1}^z) \nonumber \\
 & +\sum_{k=1}^{N/2}(\sigma_{2k}^+\sigma_{2k-1}^-\sigma_{2k+2}^-\sigma_{2k+3}^++\sigma_{2k-1}^+\sigma_{2k}^-\sigma_{2k+3}^-\sigma_{2k+2}^+).
\label{eq:spin}
\end{align}
Here, $\sigma^z_j$, $\sigma^+_j$, and $\sigma^-_j$ are Pauli matrices acting on the $j$-th site.
In this calculation, we have used the Jordan-Wigner transformation,
\begin{align}
\sigma_i^z & =2n_i-1 \nonumber \\
\sigma_i^+ & =c_i^\dagger\prod_{j=1}^{i-1}(1-2n_j) \label{eq:JWtr} \\
\sigma_i^- & =c_i\prod_{j=1}^{i-1}(1-2n_j). \nonumber
\end{align}
In the Nicolai model, the ground-state energy is exactly zero, i.e., SUSY is unbroken. We can verify this by noting that the fully filled state,
\begin{equation}
\dots \bullet \bullet \bullet \bullet \bullet \bullet \bullet \bullet \bullet \bullet \bullet \bullet \bullet \bullet \bullet \bullet \dots \nonumber
\end{equation}
is annihilated by both $Q$ and $Q^\dagger$, where $\bullet$ denotes an occupied site by a spinless fermion. There are also other ground states. For example, a state in which every fourth site is empty,
\begin{equation}
\dots \circ \bullet \bullet \bullet \circ \bullet \bullet \bullet \circ \bullet \bullet \bullet \circ \bullet \bullet \bullet \dots \nonumber
\end{equation}
is another ground state. Here, $\circ$ denotes an empty site. In the Nicolai model, the number of the ground states degeneracy grows exponentially with systems size. The classification of the ground states which are product states is discussed \cite{Katsura_2020}, and the number of the ground states of the Nicolai model is calculated using homology argument \cite{La_2018}. A part of the degeneracy can be understood from the first term in Eq.~(\ref{eq:spin}). The term can be interpreted as the frustrated Ising model on a one-dimensional $\Delta$ chain, which is shown in Fig. \ref{fig:deltachain}. It is instructive to consider the product states annihilated by both $Q$ and $Q^\dagger$, which are the ground states of the first term in Eq.~(\ref{eq:spin}). Using the Jordan-Wigner transformation Eq. (\ref{eq:JWtr}), $\bullet$ is mapped to an up-spin, while $\circ$ is mapped to a down-spin. The ground states of the Ising Hamiltonian for each local triangle are the following six state: $\{$\mbox{$\bullet \bullet \bullet$} , \mbox{$\circ \bullet \bullet$}, \mbox{$\bullet \bullet \circ$}, \mbox{$\circ \circ \circ$}, \mbox{$\bullet \circ \circ$}, \mbox{$\circ \circ \bullet$}$\}$. Therefore, a product state in which the spin configuration of any triangle is one of the six configurations is a ground state of the first term in Eq.~(\ref{eq:spin}). One can also check that these states are annihilated by the second term in Eq.~(\ref{eq:spin}). This explains why the ground-state degeneracy of the Nicolai model grows exponentially with the system size. Since this frustration often appears in SUSY QM lattice models, this is called {\it superfrustration} \cite{PRL_Fendley_2005, PRL_Huijse, EPJB_huijse, NJP_Huijse}.

The relation between spontaneous SUSY breaking and NG fermions is discussed by extending the Nicolai model. The author and collaborators introduced the extended Nicolai model~\cite{U1Nicolai} to study spontaneous SUSY breaking and NG fermions in nonrelativistic systems. The supercharge of the model is given by
\begin{align}
Q=\sum_{l=1}^{N/2}gc_{2l-1}+c_{2l-1}c_{2l}^\dagger c_{2l+1}.
\end{align}
Here, $g$, $N$ are a real number and the system size, respectively. This supercharge is the same as one of the Nicolai model when $g = 0 $. In this sense, this model is one parameter extension of the Nicolai model. The Hamiltonian is defined as $H=\{Q^\dagger, Q\}$. This model describes interacting spinless fermions in one-dimensional lattice. In this model, SUSY is broken spontaneously when $g\neq 0$ in both finite and the infinite systems~\cite{PRD_Moriya}. When SUSY is spontaneously broken, there exist gapless modes. Using the exact diagonalization, we show the dispersion relation of the gapless modes is linear in momentum. To obtain a further result, we also calculate the central charge $c$ of conformal field theory and the Tomonaga-Luttinger parameter $K$. Employing the exact diagonalization, we find that the lowest-lying states are described by $c=1$ CFT and the NG fermions are massless Thirring fermions with $K\sim 1$. 

The author and collaborators also introduced another model called $\mathbb{Z}_2$ Nicolai model~\cite{Z2Nicolai}. The supercharge of the model is 
defined as 
\begin{align}
Q=\sum_{j=1}^Ngc_j+c_{j-1}c_jc_{j+1}.
\end{align}
Here, $g$ and $N$ are a real number and the system size, respectively. The Hamiltonian is given by $H=\{Q^\dagger, Q\}$. In this model, the fermion number is not conserved but fermionic parity is conserved. In this sense, this model has $\mathbb{Z}_2$ symmetry. The model describes interacting spinless fermion in one-dimensional lattice. Similar to the Nicolai model, this model shows superfrustration when $g=0$, i.e., the number of the ground states degeneracy grows exponentially with system size $N$. The number of degenerated ground states is calculated using homology argument~\cite{La_2018}. In this model, SUSY is spontaneously broken when $g = 0$ in finite systems and $g>4/\pi$ in the infinite volume limit. There exist gapless modes associated with spontaneous SUSY breaking in this model. Using the exact diagonalization method, we find that the dispersion relation of NG fermions is cubic in momentum.
Through these models, we see that there exist NG fermions associated with spontaneous $\mathcal{N}=2$ SUSY breaking in non-relativistic systems.

Even though we unveiled the properties of spontaneous SUSY breaking and NG fermions in non-relativistic systems, few models with SUSY are discussed. Hence, we introduce other models with $\mathcal{N}=1$ SUSY in this thesis.


\section{Majorana fermion}
\label{majo_intro}
In this section, we would like to touch on Majorana fermions \cite{Majorana2008} introduced by E. Majorana, in the context of condensed matter physics. Majorana fermions have the property that Majorana fermions themselves are their anti-particles. In condensed matter physics, Majorana fermions are discussed in the context of topological superconductors \cite{Kitaev_2001, Read_PRB} and Kitaev materials \cite{Kitaev20062, Jackeli_PRB}. Majorana fermions in topological phases of matter attract renewed attention in terms of application to quantum information \cite{Nayak_RMP}.  The one-dimensional p-wave superconductor introduced by Kitaev is one of the most famous examples \cite{Kitaev_2001}. Kitaev studied the following Hamiltonian
\begin{align}
H_{\rm kit} = \sum_{j=1}^{L-1}\left\{-t(c^\dagger_{j+1}c_j+c^\dagger_jc_{j+1})+\Delta (c_jc_{j+1}+c^\dagger_{j+1}c^\dagger_j)\right\}-\mu \sum_{j=1}^{L}(c^\dagger_jc_j-\frac{1}{2}).
\end{align}
Here, $L$, $t$, $\Delta$, $\mu$ are the length of chain, hopping amplitude, pairing energy, and chemical potential, respectively. Symbols $c_j$ and $c_j^\dagger$ denote annihilation and creation operators of $j$-th site, respectively. $c_i^\dagger$ and $c_j$ satisfy the following canonical anticommutation relation, $\{c_i^\dagger,c_j\}=\delta_{i,j}$. This Hamiltonian describes a p-wave superconductor. For simplicity, we assume $t = \Delta$  in the rest of this chapter.
Introducing Majorana fermion operators, $H_{\rm kit}$ can be rewritten as 
\begin{align}
H_{\rm kit} ={\rm i}t \sum_{j=1}^{L-1}b_ja_{j+1} -\frac{{\rm i}\mu}{2}\sum_{j=1}^{L}a_jb_j.
\end{align}
Here, $a_j = c_j + c_j^\dagger$ and $b_j ={ \rm i}(c_j^\dagger-c_j)$. These $a_j$ and $b_j$ satisfy the following relation
\begin{align*}
a_j^\dagger = a_j \quad , \quad b_j^\dagger = b_j,
\end{align*}
which indicates the property that Majorana fermion itself is anti-particle.
In the case of $t = 0$, the Kitaev Hamiltonian leads
\begin{align}
H_{\rm kit} = -\frac{{\rm i}\mu}{2}\sum_{j=1}^{L}(a_jb_j).
\end{align}
In this case, the ground state $|\Phi_0\ra$ is unique and called trivial phase.
In the case of $\mu = 0$, the Kitaev Hamiltonian reads
\begin{align}
H_{\rm kit} ={\rm i}t \sum_{j=1}^{L-1}(b_ja_{j+1}).
\end{align}
In this case, the ground states are doubly degenerate since there are unpaired two Majorana fermions at the edges. Let $|\Phi_1\rangle$ be the ground state of the model, $|\Phi_1\rangle$ and $\frac{1}{2}(a_1 -b_L)|\Phi_1\rangle$ are doubly degenerated and called topological phase.
In nature, there are interactions between particles. Sometimes such interactions give interesting phenomena such as reduction of $\mathbb{Z}$ symmetry of topological classification of free fermions to $\mathbb{Z}_8$ symmetry\cite{PRB_Fidkowski_10, PRB_Fidkowski_11}. In Majorana Hubbard models, interactions give rich phase diagrams including emergent SUSY \cite{Rahmani_PRB, Milsted_PRB, Affleck_PRB}. 
\begin{figure}[htb]
\begin{center}
\includegraphics[width=0.8\columnwidth]{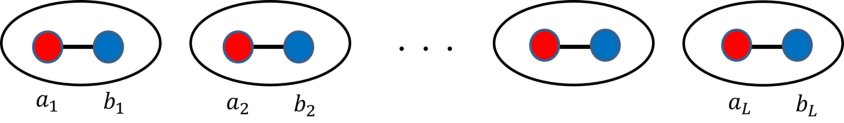}
\end{center}
\caption{ The graphical representation of the Kitaev Hamiltonian $H_{\rm kit}$ in the trivial phase. Red circles and blue circles denote $a_j$ and $b_j$, respectively. The horizontal line between Majorana fermions denotes hopping of the Majorana fermion.}
\label{fig:chap1_triv_phase}
\end{figure}
\begin{figure}[htb]
\begin{center}
\includegraphics[width=0.8\columnwidth]{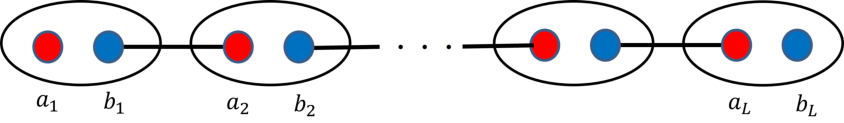}
\end{center}
\caption{ The graphical representation of the Kitaev Hamiltonian $H_{\rm kit}$ in the topological phase. Red and blue circles denote $a_j$ and $b_j$, respectively. The horizontal line between Majorana fermions denotes hopping of the Majorana fermion.}
\label{fig:chap1_topo_phase}
\end{figure}

\section{Organization}
This thesis is organized as follows.

In Chap.~\ref{chap:N1susy}, we first provide superalgebra of $\mathcal{N}=1$ SUSY and discuss the general properties of $\mathcal{N} = 1$ SUSY. We next give a definition of SUSY breaking.

In Chap.~\ref{chap:MajoNicolai}, we introduce the Majorana-Nicolai model with one parameter $g$. In Sec.~\ref{sec:chap3_susy}, we first prove that SUSY is unbroken and the ground states are solvable when $g = 1$. Following this, we find that SUSY is restored in an extended area of $g$. For $g>8/\pi-1$, we prove that SUSY is spontaneously broken by giving a lower bound of the ground-state energy density. In Sec.~\ref{sec:chap3_NGmodes}, we show that spontaneous SUSY breaking gives rise to gapless modes by giving an upper bound of excitation energy. In Sec.~\ref{sec:chap3_dispersion}, we find that the dispersion relation of NG modes is cubic in momentum.

In Chap.~\ref{chap:LNicolai}, we introduce another model with SUSY which describes interacting Majorana fermions. This model has one parameter $g$. In Sec.~\ref{sec:chap4_susy}, we first prove SUSY is unbroken in the case of $g = 1$. Next, we find that SUSY is restored in the infinite volume limit. For $g>8/\pi-1$, we prove that SUSY is spontaneously broken. In Sec.~\ref{sec:chap4_NGmodes}, we show that there exist gapless modes associated with spontaneous SUSY breaking. In Sec.~\ref{sec:chap4_dispersion}, we find that the dispersion relation of NG modes is linear in momentum and the lowest-lying states can be described by $c = 1/2$ conformal field theory.

In Chapter~\ref{sec:conclusion}, we provide a summary of this thesis. In Appendices ~\ref{chap:appa} and~\ref{chap:appb}, we provide some formulas and calculations needed  in the main text.

\chapter{$\mathcal{N} = 1$ SUSY Quantum Mechanics} 

\label{chap:N1susy} 

In this chapter, we first introduce $\mathcal{N} = 1$ SUSY Quantum Mechanics (QM) and derive general properties of $\mathcal{N} = 1$ SUSY QM. Next, we give the definition of spontaneous SUSY breaking.



\section{$\mathcal{N} = 1$ SUSY algebra and its properties}
\label{sec:algebra}
In this section, we introduce $\mathcal{N} = 1$ SUSY algebra and its general properties. Here, we give a definition of $\mathcal{N} = 1$ SUSY as
\begin{align*}
H = Q^2, \quad [H, Q] = 0, \quad Q^\dagger = Q, \quad \{Q, (-1)^F\} = 0 
\end{align*}
where $H$ is the Hamiltonian of the system, $Q$ is the supercharge, and $(-1)^F$ is the fermionic parity. Since the supercharge $Q$ anti-commutes with the fermionic parity $(-1)^F$ and commutes with the Hamiltonian $H$, $Q$ exchanges bosonic states and fermionic states with the same energy. Here we refer to bosonic states and fermionic states as states with $(-1)^F=1$ and $(-1)^F = -1$, respectively. The Hamiltonian is given by the square of the supercharge $Q$, and this leads to that $Q$ is a conserved quantity. The Hamiltonian is positive-semidefinite
\begin{align}
\langle \psi | H |\psi\rangle = \langle\psi | Q^2|\psi\rangle = \|Q|\psi\rangle\|^2 \geq 0.
\end{align}
Here, $|\psi\rangle$ is an arbitrary state. Therefore, $|\psi\rangle$ must be a zero-energy ground state when it is anihilated by $Q$ i. e., $Q|\psi\rangle \ = \ 0$.

\begin{figure}[htb]
\begin{center}
\includegraphics[width=0.8\columnwidth]{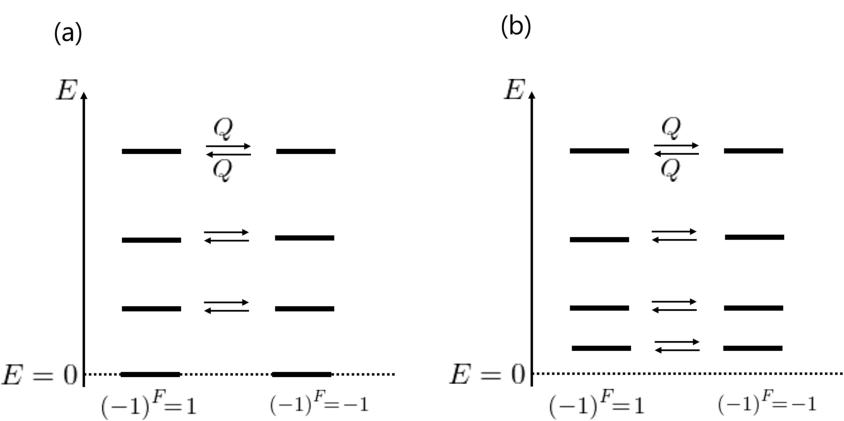}
\end{center}
\caption{ (a) The schematics of energy spectra in the case of unbroken SUSY. The ground states are zero-energy states. Excited states with opposite fermionic parity are doubly degenerate because of SUSY. (b) The schematics of energy spectra in the case of broken SUSY. Ground states have positive energy and each state with opposite fermionic parity is doubly degenerate because of SUSY.}
\label{fig:chap2_energy}
\end{figure}

${\mathcal N} = 1$ SUSY SYK model \cite{PRD_Fu} is an example of a model with $\mathcal{N} = 1$ SUSY in condensed matter physics. This model describes a $0$-dimensional interacting model and is studied in terms of the relation between quantum chaos and black hole. This model shows ${\mathbb Z}_8$ periodicity of the ground-state energy similar to the breakdown of ${\mathbb Z}$ symmetry because of interaction \cite{PRB_Fidkowski_10, PRB_Fidkowski_11}. Another example is emergent SUSY in topological phases of matter \cite{Science_Grover, PRL_OBrien}. In this case, even though a system does not have SUSY, at the critical point, the low-lying states can be described using SUSY.

\section{Spontanous SUSY Breaking}
\label{sec:SSB}
In this section, we would like to introduce a definition of spontaneous SUSY breaking. In finite systems, let $|{\rm GS}\rangle$ be the ground state, spontaneous SUSY breaking can be described as
\begin{align*}
Q|{\rm GS}\rangle \neq 0 .
\end{align*}
This condition is equivalent to the condition that the ground state energy is positive since the Hamiltonian is given by the square of $Q$. In this sense, the energy is an order parameter to determine whether SUSY is broken or not. The condition is not suitable for the definition of spontaneous SUSY breaking since it contains a subtle point. In many cases, the energy increases as the volume of the system increases. However, sometimes there are exceptions. Witten pointed that SUSY may be restored in the infinite volume limit even though SUSY is broken spontaneously in finite systems. In order to avoid this subtle point, we give a strict definition of SUSY breaking. \\

\textbf{ Definition} \\
\textit{ SUSY is said to be spontaneously broken when the ground state energy density is positive.}
\\
This definition plays an important role to prove that there exist gapless modes associated with spontaneous SUSY breaking in Sec.~\ref{sec:chap3_NGmodes} and Sec.~\ref{sec:chap4_NGmodes}. In this sense, the positivity of the ground state energy density is more important than the positivity of the ground state energy itself.

We also note that the Witten index is useful to study spontaneous SUSY breaking. The Witten index is defined as
\begin{align}
\mathcal{W}={\rm Tr}[(-1)^Fe^{-\beta H}].
\end{align}
Here, $\beta$ is the inverse temperature. The Witten index plays an important role in determining whether SUSY is broken or unbroken. When energy is positive, fermions and bosons with the same energy must come in pairs. Therefore, their contributions to $\mathcal{W}$ cancel out, yielding 
\begin{align}
\mathcal{W}=n^b_0-n^f_0,
\end{align}
where $n^b_0$ and $n^f_0$ are the number of bosonic states with zero energy and that of fermionic states, respectively. According to this nature,  the Witten index gives the lower bound of the number of states with zero energy. Here, we note that the Witten index sometimes becomes zero even if SUSY is unbroken.


\chapter{Majorana Nicolai model} 

\label{chap:MajoNicolai} 

In this chapter, we introduce and study an interacting Majorana model with $\mathcal{N} = 1$ SUSY. The model has one parameter $g$ which connects the strongly interacting limit and the free limit. In Sec.~\ref{sec:chap3_model}, we give the supercharge and the Hamiltonian. In Sec.~\ref{sec:chap3_susy}, we discuss the properties of SUSY depending on the parameter $g$. In Sec.~\ref{sec:chap3_NGmodes}, we prove that there exist gapless modes when SUSY is broken spontaneously using the variational method.  In Sec~\ref{sec:chap3_dispersion}, using a numerical method, we show the dispersion relation is cubic in momentum. This chapter is based on a published paper~\cite{MajoNicolai}.


\section{Model}
\label{sec:chap3_model}
The supercharge of the model is defined as
\begin{align}
Q = \sum_{j=1}^N g\gamma_{j} + {\mathrm i}\gamma_{j-1}\gamma_{j}\gamma_{j+1},
\end{align}
where $N$ and $g$ are a total number of sites and a real parameter, respectively. Here, $\gamma_j$ is a Majorana fermion operator of $j$-th site,  a hermitian operator ($\gamma_j^\dagger = \gamma_j$), and satisfies the following anti-commutation relation (Clifford algebra) : $\{\gamma_j, \gamma_i\} = 2\delta_{i,j}$. We assume that $N$ is an even number since the dimension of the Hilbert space is $2^{N/2}$ and should be an integer. We also assume periodic boundary conditions (PBC), i.e., $\gamma_{j + N} = \gamma_j$. The fermionic parity $(-1)^F$ is defined in terms of $\gamma_j$ as $(-1)^F={\mathrm i}^{N/2}\prod_{j = 1}^N\gamma_j$.

\begin{figure}[htb]
\begin{center}
\includegraphics[width=0.8\columnwidth]{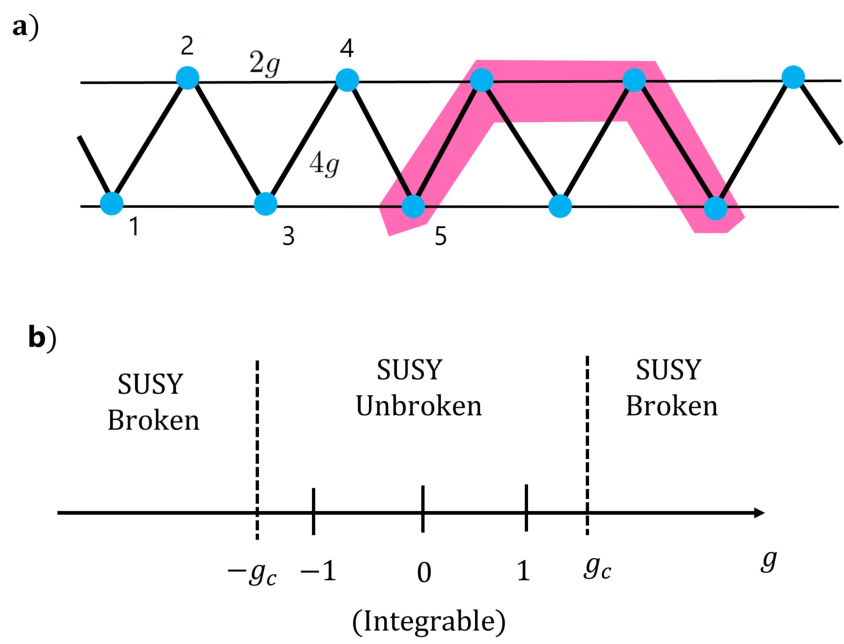}
\end{center}
\caption{ (a) The schematic graph of the Hamiltonian.  (b) The schematic phase diagram. In the figure (a), the blue dots represent sites. $4g$ and $2g$ represent hopping amplitudes of nearest neighbors and next nearest neighbors, respectively. The pink area represents interaction of Majorana fermions. The figure (b) represents the phase diagram in the infinite volume limit. When $|g|$ is large enough, SUSY is spontaneously broken. When $|g|$ is small enough, SUSY is unbroken. In the case of $g =\pm1$, the ground states are solvable. In the case of $g = 0$, the model is integrable \cite{JPA_Fendley}.}
\label{fig:chap3_schematics}
\end{figure}

The Hamiltonian is defined by the square of $Q$ and can be decomposed into three parts as follows
\begin{align}
H = Ng^2 + H_{\rm free} + H_{\rm int}, \label{eq:chap3_Ham}
\end{align}
where
\begin{align}
H_{\rm free} = 2g\mathrm{i}\sum_{j = 1}^N\left(2\gamma_j\gamma_{j+1}-\gamma_{j-1}\gamma_{j+1}\right),
\end{align}
\begin{align}
H_{\rm int} = \sum_{j = 1}^N\left(1-2\gamma_{j-1}\gamma_{j}\gamma_{j+2}\gamma_{j+3}\right).
\end{align}
Here, the first part is a constant term. The second term $H_{\rm free}$ describes the hopping of Majorana fermions between nearest neighbors and next-nearest neighbors as shown in Fig. \ref{fig:chap3_schematics} (a). The last term $H_{\rm int}$ describes the interaction of Majorana fermions in pink area in Fig. \ref{fig:chap3_schematics} (a). We note that $H_{\rm int}$ is positive semi-definite since it is given by the square of $Q$ in the case of $g = 0$.

The model has translational symmetry by one site of Majorana fermion
\begin{align*}
T^{-1}HT = H.
\end{align*}
Here, $T$ is the translational operator of Majorana fermion by one site, i.e., $T^{-1}\gamma_jT = \gamma_{j+1}$. The fermionic parity $(-1)^F$ and the translational operator $T$ anti-commute each other. One can show this as follows
\begin{align*}
T^{-1}(-1)^FT & = {\mathrm i}^{N/2}\prod_{j = 1}^N\gamma_{j+1} = {\mathrm i}^{N/2}\prod_{j = 2}^N\gamma_{j} \gamma_{N + 1} = (-1)^{N-1} \ {\mathrm i}^{N/2}\gamma_{N + 1}\prod_{j = 2}^N\gamma_{j} \\
& = (-1)^{N-1} \ {\mathrm i}^{N/2}\gamma_{1}\prod_{j = 2}^N\gamma_{j} =-{\mathrm i}^{N/2}\prod_{j = 1}^N\gamma_{j}, 
\end{align*}
where we use the fact that $N$ is an even number.

\bigskip

\section{Properties of SUSY}
\label{sec:chap3_susy}
In this section, we discuss the properties of the model in terms of SUSY breaking.
\subsection{Unbroken SUSY \label{sec:unbroken}}
In this subsection, we prove that SUSY is unbroken and the ground state is solvable when $g = 1$. For the latter purpose, we introduce the Hamiltonian of Kitaev chain with PBC as follows
\begin{align*}
H_{\rm kit} = \sum_{j}^L\left(-t(c_j^\dagger c_{j+1}+c_{j+1}^\dagger c_j) -\mu(c^\dagger_jc_j - \frac{1}{2})+\Delta (c_jc_{j+1}+c_{j+1}^\dagger c_j^\dagger)\right)
\end{align*}
where $L$, $t$, $\mu$, and $\Delta$ are system size, hopping amplitude, on site potential and pairing potential, respectively. Here, $c_j$ and $c_j^\dagger$ are annihilation operator of fermion on $j$-th site and creation operator of fermion on $j$-th site. By rewriting the Kitaev Hamiltonian using Majorana fermions, we obtain
\begin{align}
H_{\rm kit} = \frac{\mathrm{i}}{2}\sum_{j=1}^L(-\mu\gamma_{2j-1}\gamma_{2j}+(t+\Delta)\gamma_{2j}\gamma_{2j+1}+(-t+\Delta)\gamma_{2j-1}\gamma_{2j+2}),
\end{align}
where $\gamma_{2j-1}=c_j + c_j^\dagger$ and $\gamma_{2j}={\mathrm i}(c^\dagger_j-c_j)$.
Here, we consider the following two specific cases of the Kitaev chain Hamiltonian
\begin{align*}
H_{\rm kit} =-\frac{\mathrm{i}\mu}{2}\sum_{j=1}^L\gamma_{_{2j-1}}\gamma_{_{2j}} \quad (t = \Delta = 0, \mu < 0)
\end{align*}
and
\begin{align*}
H_{\rm kit}=\mathrm{i}t\sum_{j=1}^L\gamma_{2j}\gamma_{2j+1} \quad (t = \Delta \neq 0, \mu = 0).
\end{align*}
In the case of $\mu < 0$ and $t = \Delta = 0$, the ground state of the Kitaev chain $|\Phi_0\rangle$ is annihilated by local operators $(1 + \mathrm{i}\gamma_{2l-1}\gamma_{2l})$ as,
\begin{align*}
(1 + \mathrm{i}\gamma_{2l-1}\gamma_{2l})|\Phi_0\rangle = 0.
\end{align*}
In the case of $t = \Delta\neq 0$ and $\mu = 0$, the ground state of the Kitaev chan $|\Phi_1\rangle$ is annihilated by local operators $(1 + \mathrm{i}\gamma_{2l}\gamma_{2l+1})$ as,
\begin{align*}
(1 + \mathrm{i}\gamma_{2l}\gamma_{2l+1})|\Phi_1\rangle = 0.
\end{align*}
These two ground states are also the ground states of our model. To show that $|\Phi_0\rangle$ and $|\Phi_1\rangle$ are ground states of our model, we rewrite supercharge $Q$ as follows
\begin{align}
Q & = \sum_{l = 1}^{N/2}(\gamma_{2l - 2}+\gamma_{2l+1})(1 + \mathrm{i}\gamma_{2l - 1}\gamma_{2l}) \\
& = \sum_{l = 1}^{N/2}(\gamma_{2l-1} + \gamma_{2l+2})(1 + \mathrm{i}\gamma_{2l}\gamma_{2l+1}).
\end{align}
Since $Q$ contains $(1 + \mathrm{i}\gamma_{2l - 1}\gamma_{2l})$ and $(1 + \mathrm{i}\gamma_{2l}\gamma_{2l+1})$, $Q$ annihilates $|\Phi_0\rangle$ and $|\Phi_1\rangle$
\begin{align*}
Q|\Phi_0\rangle=Q|\Phi_1\rangle=0.
\end{align*}
Therefore, SUSY is unbroken in the infinite system. Since the Hamiltonian is positive semidefinite, $|\Phi_0\rangle$ and $|\Phi_1\rangle$ must be the ground states of the model. We also calculate the number of the ground state degeneracy depending on $N$. The results of exact diagonalization are shown in Table~\ref{tab:gs}. From this table, we can see that there are other two ground states when $N$ is a multiple of $8$. 
\begin{table}
\begin{tabular}{c|ccccc}\hline
~~$N \ {\rm mod}\ 8$~~ & 0 & 2 & 4 & 6\\ \hline
~~$\#GS$~~ & ~4~ & ~2~ & ~2~ & ~2~\\ \hline
\end{tabular}
\caption{The ground state degeneracy depending on $N$.}
\label{tab:gs}
\end{table}
The other ground states are
\begin{align*}
|\Psi_0\rangle=\frac{1}{\sqrt{N_0}}\sum_{j=1}^Ne^{-{\rm i}\frac{\pi}{4}j}\gamma_j|\Phi_0\rangle
\end{align*}
and
\begin{align*}
|\Psi_1\rangle = T|\Psi_0\rangle.
\end{align*}
Here, $N_0$ is a normalization coefficient. We discuss that $|\Psi_0\rangle$ and $|\Psi_1\rangle$ are ground states of the model in Appendix~\ref{appA:dgs}.

\subsection{Restoration of SUSY\label{sec:restore}}
\begin{figure}[htb]
\begin{center}
\includegraphics[width=0.8\columnwidth]{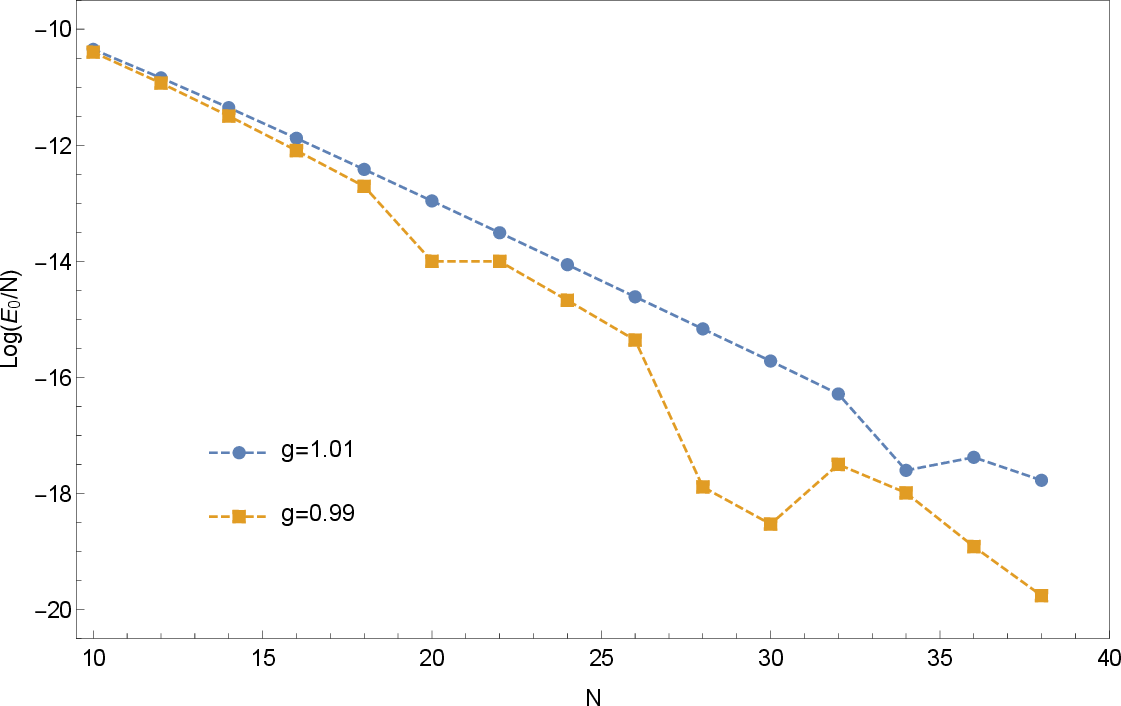}
\end{center}
\caption{The log-linear plot of the ground state energy density for $g = 0.99$ and $1.01$.}
\label{fig:chap3_gden-all}
\end{figure}

In this subsection, we discuss restoration of SUSY. 
When $g$ is close to $1$, we find that SUSY is restored in the thermodynamic limit. The Fig.~\ref{fig:chap3_gden-all} shows the results of exact diagonalization for $g = 0.99$ and $g = 1.01$ with $N = 10, \dots, 38$. The vertical axis represents the logarithm of the ground state energy density, and the horizontal axis represents $N$. 
From Fig.~\ref{fig:chap3_gden-all}, we can see that the ground state energy density decreases exponentially as $N$ increases. Therefore, the ground state energy density is likely to go to zero in the thermodynamic limit so that SUSY is considered to be restored in the thermodynamic limit.

\begin{figure}[htb]
\begin{center}
\includegraphics[width=0.8\columnwidth]{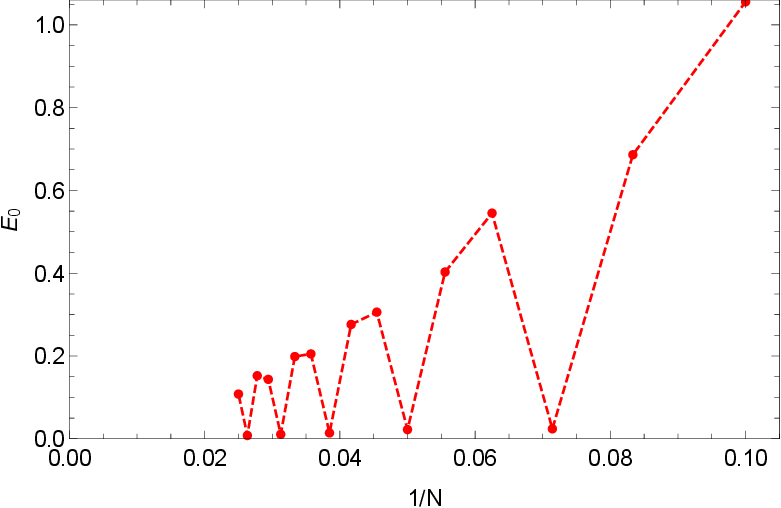}
\end{center}
\caption{The ground state energy for $g = 0$ depending on $N$.}
\label{fig:chap3_gene0}
\end{figure}

In the case of $g = 0$, we also find that SUSY is considered to be restored in the thermodynamic limit. In this case, we have calculated the ground state energy. The Fig.~\ref{fig:chap3_gene0} shows the results of exact diagonalization with $N = 10, \dots, 38$. The vertical axis represents the ground state energy, and the horizontal axis represents $1 / N$.
From Fig.~\ref{fig:chap3_gene0}, we can see that the ground state energy has a $6$-fold periodicity depending on $N$. We note that the supersymmetric Sachdev-Ye-Kitaev model has $8$-fold periodicity depending on $N$~\cite{PRD_Fu}. From Fig.~\ref{fig:chap3_gene0}, we find that the ground state energy decreases as $N$ increases, and it is considered to converge to a certain value. Hence, the ground state energy density converges to zero in the thermodynamic limit, so that SUSY is restored in the thermodynamic limit. We note that the model is solvable when $g = 0 $\cite{JPA_Fendley}.

\subsection{SUSY breaking\label{sec:chap3_broken}}
In this subsection, we give the condition that SUSY is broken spontaneously. In the case $g>8/\pi-1$, we prove that SUSY is broken spontaneously in both finite and infinite systems. In the case of $g = 1 + \delta$, the Hamiltonian can be written as 
\begin{align}
H = N(\delta^2 + 2\delta) + 2\delta\mathrm{i}\sum_{j = 1}^N\left(2\gamma_j\gamma_{j+1}-\gamma_{j-1}\gamma_{j+1}\right)+H_{g=1}.
\label{eq:chap3_deltaexp}
\end{align} 
Here, $H_{g=1}$ is the Hamiltonian of the $g = 1$ case.
By employing Anderson's argument~\cite{PR_Anderson, PRB_Valenti, PRD_Beccaria, PRL_Nie}, we obtain the following inequality
\begin{align}
E_0 \ge N(\delta^2 + 2\delta) + E_0^{\rm free}
\label{eq:chap3_eineq}
\end{align}
where $E_0$ is the ground state energy and  $E_0^{\rm free}$ is the ground state energy of the second term in Eq. (\ref{eq:chap3_deltaexp}). Here, we use the condition that $H_{g=1}$ is positive semidefinite. Since the second term in Eq. (\ref{eq:chap3_deltaexp}) is quadratic in Majorana fermion operators, we can calculate $E_0^{\rm free}$ exactly as follows
\begin{align}
E_0^{\rm free} = -\frac{8\delta}{\tan\left(\pi/N\right)}.
\label{eq:chap3_efree}
\end{align}
The derivation of Eq.(\ref{eq:chap3_efree}) is shown in Appendix~\ref{appA:free}. From Eqs.(\ref{eq:chap3_eineq}) and (\ref{eq:chap3_efree}), we get the following inequality that gives a lower bound of the ground state energy density
\begin{align}
\frac{E_0}{N} \ge  N(\delta^2 + 2\delta)  + \frac{E_0^{\rm free}}{N} \ge \delta(\delta + 2 - \frac{8}{ \pi}).
\label{eq:chap3_gedineq}
\end{align}
From Eq. (\ref{eq:chap3_gedineq}), we can see that the ground state energy density must be positive when $\delta > 8/\pi-2$. Hence, SUSY is broken spontaneously when $g = 1+\delta > 8/\pi -1$. We note that the condition is not optimal. In Appendix~\ref{appA:ged}, we discuss that the ground state energy density can be positive even when $g \le 8/\pi-1=1.546\dots$.

\bigskip

\section{Nambu-Goldstone modes}
\label{sec:chap3_NGmodes}
The Nambu-Goldstone theorem says that a breakdown of a global and continuous symmetry gives rise to low lying excitations such that $\lim_{p\rightarrow0}E_p \rightarrow 0$. In this section, we prove the existence of gapless mode associated with spontaneous SUSY breaking. In order to prove this, we use the variational method called the Bijl-Feynman ansatz \cite{PR_Feynman} which is also used in the context of the Heisenberg antiferromagnets \cite{ZPB_Horsch, PRB_Stringari, JPSJ_Momoi}. As an example, we briefly review the dispersion relation of the Heisenberg model in $D$-dismensional lattice, where we assume $D>1$. The Hamiltonian is defined as 
\begin{align}
H=J\sum_{\langle i,j\rangle}\hat{\bm{S}}_i\cdot\hat{\bm{S}}_{j}
\end{align}
where $J$ is coupling constant, the summation is taken for nearest neighbors, and $\hat{\bm{S}}_i = (S_i^x, S_i^y, S_i^z)$ is the spin operator of $i$-th site. In this model, it is well-known that spin rotational symmetry is spontaneously broken and there must exist gapless modes. We discuss the gapless modes of this model for both ferromagnetic ($J<0$) and anti-ferromagnetic ($J>0$) cases. For the ferromagnetic case, we obtain quadratic dispersion relation, i.e. $\omega(p)\propto p^2$, by applying spin-wave approximation to all-up state, i.e., the ground state.  For the anti-ferromagnetic case, we obtain linear dispersion relation, i.e. $\omega(p)\propto |p|$, by applying spin-wave approximation to the N{\'e}el state. This approach has a subtle point for the anti-ferromagnetic case since the N{\'e}el state is not an eigenstate of the Hamiltonian. To avoid this subtlety, we use a variational method. Using the ground state $|\Psi_g\rangle$ of the Heisenberg model, we introduce a trial state $|\Psi(p)\rangle$ defined as 
\begin{align}
|\Psi(p)\rangle=S_p^z|\Psi_g\rangle/ \parallel S_p^z|\Psi_g\rangle \parallel
\end{align}
Here, $S_p^z$ is the Fourier transform of $S_i^z$ defined as $S_p^z=\frac{1}{\sqrt N}\sum_i S_i^ze^{{\mathrm{i}p\cdot r_i}}$ where $N$ is a normalization coefficient. We introduce a variational energy $\epsilon(p)$ defined as
\begin{align*}
\epsilon(p)=\langle\Psi(p)|H|\Psi(p)\rangle-\langle\Psi_g|H|\Psi_g\rangle).
\end{align*}
In the thermodynamic limit, this variational energy is bounded by $p$-linear from above as 
\begin{align*}
\epsilon(p)\le C|p|+\mathcal{O}(p^2)
\end{align*}
where $C$ is not zero. However we prove that there exist gapless modes, we cannot determine the dispersion relation of the lowest-lying states using this method. We also apply this variational method to the SUSY model. Here, we assume that $g$ is greater than $8/\pi-1$, that is SUSY is broken spontaneously, and the ground state degeneracy is of the order of $1$. Indeed, exact diagonalization calculation shows that the ground states are four-fold degenerated. We introduce the Fourier transform of local supercharge $q_j$ as follows,
\begin{align}
Q_p = \sum_{j=1}^Nq_j \cos(pj) \quad , \quad q_j = g \gamma_j + \mathrm{i}\gamma_{j-1}\gamma_{j}\gamma_{j-1}.
\end{align}
Here, $q_j$ satisfies the following locality conditions
\begin{equation}
\{q_j,q_i\} = 
\begin{cases}
{\rm nonzero} & (|j - i| \le 2) \\
0 & ({\rm otherwise}).
\end{cases}
\label{eq:local}
\end{equation}
We note that $Q_p$ is an even function of $p$, and hermitian, namely, 
\begin{align}
Q_{-p} = Q_{p} \quad , \quad Q_p^\dagger = Q_p.
\end{align}
We define a trial state as $|\psi(p)\rangle = Q_p|\psi_0\rangle$, where $|\psi_0\rangle$ is a ground state of the model. Here, I note that numerical calculation shows that the ground states is four-fold degenerate in SUSY broken region, irrespective of the system size. Since $Q_p|\psi_0\rangle$ is a linear combination of states with momenta $\pm p$, $Q_p|\psi_0\rangle$ is orthogonal to $|\psi_0\rangle$ and $Q|\psi_0\rangle$. This means that $Q_p|\psi_0\rangle$ contains information of excited states. We define a variational energy $\epsilon_{\rm var}$ using the trial state $|\psi(p)\rangle$ as follows
\begin{align}
\epsilon_{\rm var} = \frac{\langle\psi(p)|H|\psi(p)\rangle}{\langle\psi(p)|\psi(p)\rangle} - E_0.
\end{align}
Here, $H$ and $E_0$ are the Hamiltonian and the ground state energy, respectively. Since $Q_p|\psi_0\rangle$ is a linear combination of excited states, the variational energy $\epsilon_{\rm var}$ is equal to or greater than the first excitation energy. The variational energy can be rewritten using a double commutator as follows,
\begin{align}
\epsilon_{\rm var} & = \frac{\langle\psi(p)|H|\psi(p)\rangle}{\langle\psi(p)|\psi(p)\rangle} - E_0 \nonumber \\
& = \frac{\langle\psi_0|[Q_p, [H, Q_p]]|\psi_0\rangle}{2\langle\psi(p)|\psi(p)\rangle} \nonumber \\
& = \frac{\langle\psi_0|[Q_p, [H, Q_p]]|\psi_0\rangle}{\langle\psi_0|\{Q_p, Q_p\}|\psi_0\rangle} = \frac{\langle[Q_p, [H, Q_p]]\rangle_0}{\langle\{Q_p, Q_p\}\rangle_0},
\label{eq:longeq}
\end{align}
where $\langle \cdots\rangle_0$ denotes the expectation value in the ground state. In the last line of Eq. (\ref{eq:longeq}), we use the identity $2Q_p^2 = \{Q_p, Q_p\}$. To obtain the upper bound, we introduce the Pitaevskii-Stringali inequality~\cite{JLTP_Pitaevskii},
\begin{align}
|\langle\phi|[A^\dagger, B]|\phi\rangle|^2\le\langle\phi|\{A^\dagger, A\}|\phi\rangle\langle\phi|\{B, B^\dagger\}|\phi\rangle \label{eq:PitaevskiiStringari}
\end{align}
where $|\phi\rangle$ is an arbitrary state, and $A$ and $B$ are any operators. Using this inequality, we obtain the following inequality
\begin{align}
(\epsilon_{\rm var})^2 \le \frac{\langle\{[H,Q_p], [Q_p,H]\}\rangle_0}{\langle\{Q_p, Q_p\}\rangle_0}.
\end{align}
Since the numerator of the right hand side, $\langle\{[H,Q_p], [Q_p,H]\}\rangle_0$, is an even function of momentum $p$, it can be expanded as follows when $p$ is small enough
\begin{align}
\langle\{[H,Q_p],[Q_p,H]\}\rangle_0=NC_2p^2+\mathcal{O}(p^4).
\label{eq:numerator}
\end{align}
Here, $C_2$ is a constant coefficient, and the $p$ independent term in Eq. (\ref{eq:numerator}) must be $0$ because $Q_p$ becomes the supercharge $Q$ when $p = 0$ and $Q$ is a conserved quantity, i.e., $[H, Q] = 0$.
Since the denominator of the right hand side, $\langle\{Q_p, Q_p\}\rangle_0$, is an even function of momentum $p$, it can be expanded as follows when $p$ is small enough
\begin{align}
\langle\{Q_p, Q_p\}\rangle_0=2E_0+\mathcal{O}(p^2).
\label{eq:denominator}
\end{align}
Here, $E_0$ is the ground state energy. The right hand sides of both Eqs. (\ref{eq:numerator}) and (\ref{eq:denominator}) are of order of $N$ because of locality condition Eq. (\ref{eq:local}). From Eqs. (\ref{eq:numerator}) and (\ref{eq:denominator}), we obtain the following inequality when $p$ is small enough
\begin{align}
\epsilon_{\rm var} \le \sqrt{\frac{C_2}{E_0 / N}} |p| + \mathcal{O}(p^2). \ 
\end{align}

From this inequality, we see that there exist gapless modes when the momentum $p$ is small enough. The definition of spontaneous SUSY breaking plays an important role in this inequality since the numerator of the right-hand side contains the ground state energy density $E_0 / N$. If SUSY is not broken ($E_0 / N$ = 0), the upper bound of the inequality does not make sense.
We note that we assume only supersymmetry to prove the inequality. Therefore, we can apply this inequality for any model with $\mathcal{N} = 1$ SUSY.

\bigskip

\section{Dispersion relation}
\label{sec:chap3_dispersion}
\begin{figure}[htb]
\begin{center}
\includegraphics[width=1.0\columnwidth]{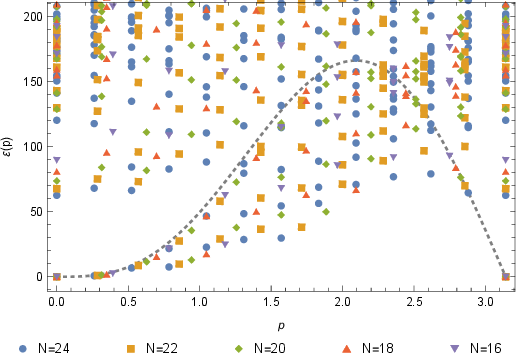}
\end{center}
\caption{Many body spectra of the Hamiltonian for $g = 8$ with $N = 16, \dots , 24$. The horizontal axis represents momentum $p$. The dotted curve is one particle excitation of $H_{\rm free}$, i.e., $8gf(p)$.}
\label{fig:chap3_disp}
\end{figure}

In the previous section, we have proved the existence of gapless modes associated with spontaneous SUSY breaking. In this section, we unveil the dispersion relation of the gapless mode in the Majorana-Nicolai model when SUSY is broken spontaneously. First of all, we consider the dispersion relation in the large-$g$ limit. In the large-$g$ limit, the interaction term $H_{\rm int}$ is negligible, yielding
\begin{align}
H \sim Ng^2 + H_{\rm free} \quad , \quad H_{\rm free} = 2g{\mathrm i}\sum_{j = 1}^N \left(2\gamma_j\gamma_{j+1} -\gamma_{j-1}\gamma_{j+1}\right).
\end{align}
Using the Fourier transform, $H_{\rm free}$ is rewritten as follows,
\begin{align}
H_{\rm free} = 8g\sum_{p > 0}f(p)\gamma(p)^\dagger\gamma(p) - \frac{8g}{\tan(\pi/N)}.
\label{eq:fourierfree}
\end{align}
Here, $f(p) = \left(2\sin(p) - \sin(2p)\right)$, $\gamma(p)$ is the Fourier transform of the Majorana fermion operators and defined as
\begin{align*}
\gamma(p) = \sqrt{\frac{1}{2N}}\sum_{j=1}^N\gamma_j\exp({\mathrm i}pj)
\end{align*}
and $\gamma(p)^\dagger$ is defined by $\gamma(-p)$, $\gamma(p)^\dagger = \gamma(-p)$. By the McLaughlin expansion, we get $f(p) = p^3 + {\mathcal O}(p^5)$. From the Eq. (\ref{eq:fourierfree}), we find that the dispersion relation in the large-$g$ limit  is cubic in momentum $p$. Therefore, we expect that dispersion relation for finite $g$ is also cubic in momentum. The derivation of Eq. (\ref{eq:fourierfree}) is shown in Appendix~\ref{appA:Fourier}.\\
In finite $g$ cases, the Fig. \ref{fig:chap3_disp} shows the results of exact diagonalization for $N = 16, \dots, 24$ with $g = 8$. The horizontal axis represents momentum $p$, and the vertical axis represents many body spectra relative to the ground state energy. The dotted curve shows the dispersion relation of one particle excitation of $H_{\rm free}$ i.e. $f(p) = \left(2\sin(p) - \sin(2p)\right)$.
In the vicinity of the $p = 0$ point, there exist energy spectra which fit to $f(p)$. Spectra below $f(p)$ are considered to be two body or many body particle excitations.

\begin{figure}[htb]
\begin{center}
\includegraphics[width=0.9\columnwidth]{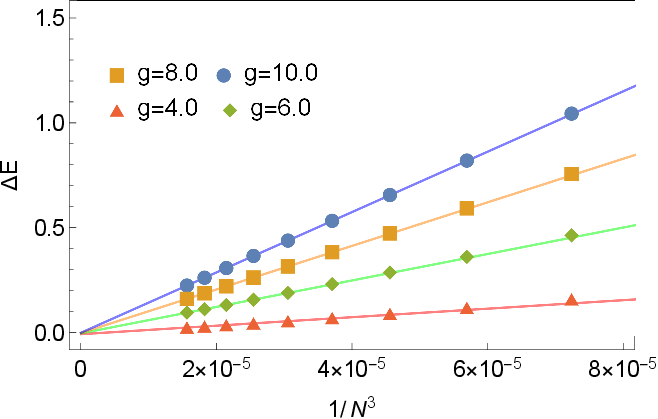}
\end{center}
\caption{The first excitation energy of the Hamiltonian for each $g$ as a function of $1/N^3$. The lines are fits to the data of $N=32, \dots,40$.}
\label{fig:first}
\end{figure}

As supporting evidence of gapless and cubic dispersion, we have calculated the first excitation energy using exact diagonalization for $g = 4, 6, 8, 10$ cases. The figure Fig. \ref{fig:first} shows the results of the calculations. From the figure, we find that the first excitation is proportional to $1/N^3$ and goes to zero in the infinte volume limit. Therefore we conclude that the dispersion relation is cubic in momentum and gapless.
In the Fig. \ref{fig:first}, lines are fits to the data of $N=32, \dots,40$. In appendix \ref{appA:notlinear}, we provide further supporting results that the dispersion relation is cubic in momentum.



\chapter{Majorana model with linear dispersion} 

\label{chap:LNicolai} 
In this chapter, we introduce and study an interacting Majorana model with $\mathcal{N} = 1$ SUSY different from one in the Chap~\ref{chap:MajoNicolai}. The model has one parameter $g$ which connects the strongly interacting limit and the free limit. In Sec.~\ref{sec:chap4_model}, we give the supercharge and the Hamiltonian. In Sec.~\ref{sec:chap4_susy}, we discuss the properties of SUSY depending on the parameter $g$. In Sec.~\ref{sec:chap4_NGmodes}, we prove that there exist gapless modes when SUSY is broken spontaneously using the variational method.  In Sec~\ref{sec:chap4_dispersion}, using a numerical method, we show that the dispersion relation is linear in momentum, and low-lying states can be described using Ising conformal field theory (CFT).


\section{Model}
\label{sec:chap4_model}
The supercharge is defined by
\begin{align}
Q = \sum_{k = 1}^{N/2}g\gamma_{2k-1} + {\mathrm i}\gamma_{2k -1}\gamma_{2k}\gamma_{2k+1}.
\end{align}
Here, $\gamma_{j}$ is a Majorana fermion operator acting on the $j$-th site, $N$ is a total number of sites, and $g$ is a real parameter. The Hamiltonian is defined as $H = Q^2$. The explicit representation of the Hamiltonian is
\begin{align}
H & = \frac{N}{2}g^2 + H_{\rm free} + H_{\rm int}.
\label{eq:chap4_Ham}
\end{align}
Here, $H_{\rm free}$ and $H_{\rm int}$ are defined as
\begin{align}
H_{\rm free} & = 2g \mathrm{i}\sum_{j = 1}^{N}\gamma_{j}\gamma_{j+1}, \\
H_{\rm int}= & -\sum_{l=1}^{N/2}(2\gamma_{2l-1}\gamma_{2l}\gamma_{2l+2}\gamma_{2l+3}-1).
\end{align}
In the rest of this chapter, we assume that $g$ is non-negative since the Hamiltonian with $g$ is mapped to the Hamiltonian with $-g$ by a unitary transformation $U$ such that $U^{-1}\gamma_jU = \gamma_{N - j}$.

\begin{figure}[htb]
\begin{center}
\includegraphics[width=1.0\columnwidth]{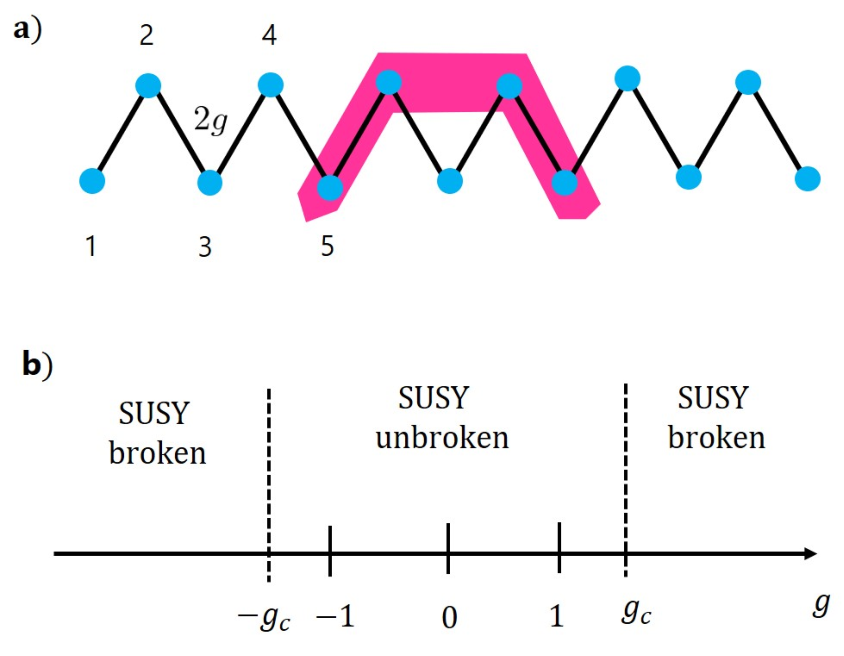}
\end{center}
\caption{(a) Schematic diagram of the Hamiltonian. The symbol $2g$ denotes the hopping amplitude of $H_{\rm free}$, and a pink area denotes interaction described by $H_{\rm int}$. (b) Schematic phase diagram in the infinite volume limit.}
\label{fig:chap4_schematic}
\end{figure}

\bigskip

\section{Properties of SUSY}
\label{sec:chap4_susy}
\subsection{Unbroken SUSY\label{sec:chap4_unbroken}}
In this subsection, we discuss the ground state when SUSY is unbroken in finite systems.
We find that SUSY is unbroken when $g = 1$, and the ground states are the same as those of the Kitaev chain with PBC. By the same strategy as the previous chapter, we rewrite the supercharge $Q$ as follows,
\begin{align*}
Q & =\sum_{l=1}^{L}\gamma_{2l+1}(1+\mathrm{i}\gamma_{2l-1}\gamma_{2l}), \\
& =\sum_{l=1}^{L}\gamma_{2l-1}(1+\mathrm{i}\gamma_{2l}\gamma_{2l+1}).
\end{align*}
Here, $L$ is defined as $L = N / 2$. Since operators $(1+\mathrm{i}\gamma_{2l-1}\gamma_{2l})$ and $(1+\mathrm{i}\gamma_{2l}\gamma_{2l+1})$ annihilate the ground states of the Kitaev chain $|\Psi_0\rangle$ and $|\Psi_1\rangle$, respectively, these two states must be the ground states of the model. Here, we mention that, by numerical calculation, the number of the ground states are two when $g = 1$. Even though the Hamiltonian does not have translational symmetry of Majorana fermion by one site, these two ground states are exchanged with each other by translation operator $T$ as follows
\begin{align}
T|\Psi_0\rangle = |\Psi_1\rangle \quad, \quad T|\Psi_1\rangle = |\Psi_0\rangle.
\end{align}
In order to show the reason, we introduce a new supercharge $\tilde{Q}$ defined as
\begin{align*}
\tilde{Q} = \sum_{l=1}^{N/2}g\gamma_{2l}+\mathrm{i}\gamma_{2l}\gamma_{2l+1}\gamma_{2l+2}.
\end{align*}
By rewriting $\tilde{Q}$, we get the following notation of $\tilde{Q}$
\begin{align}
\tilde{Q} & =\sum_{l=1}^{N/2}\gamma_{2l-2}(1+\mathrm{i}\gamma_{2l-1}\gamma_{2l}) =\sum_{l=1}^{N/2}\gamma_{2l+2}(1+\mathrm{i}\gamma_{2l}\gamma_{2l+1}).
\label{eq:tilde_Q}
\end{align}
From Eq. (\ref{eq:tilde_Q}), we can see that $\tilde{Q}$ also annihilates $|\Psi_0\rangle$ and $\Psi_1\rangle$. Since $Q$ and $\tilde{Q}$ exchanges each other by translation operator $T$
\begin{align*}
T^{-1}QT = \tilde{Q} \quad, \quad T^{-1}\tilde{Q}T = Q, 
\end{align*}
the ground states $|\Psi_0\rangle$ and $|\Psi_1\rangle$ also exchange each other by translation operator $T$.

\bigskip

\subsection{restoration of SUSY\label{sec:chap4_restore}}
In this subsection, we unveil that there is an extended area in the parameter space where SUSY is restored in the infinite volume limit. We note that numerical calculation shows that SUSY is spontaneously broken infinite systems when $g \neq 1$. 

\begin{figure}[htb]
\begin{center}
\includegraphics[width=1.0\columnwidth]{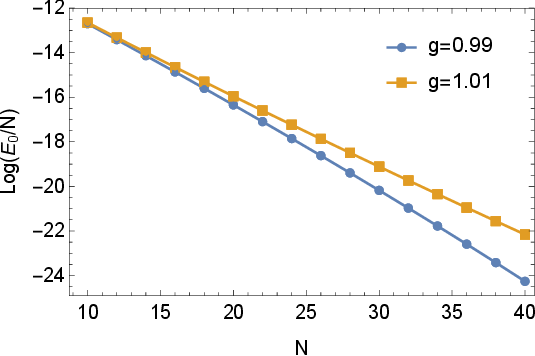}
\end{center}
\caption{The log linear plots of the ground state energy density for $g = 0.99, 1.01$.}
\label{fig:chap4_eneden}
\end{figure}
Numerical calculations show that SUSY is restored in the thermodynamic limit in the vicinity of unbroken SUSY point $g = 1$. The Fig.~\ref{fig:chap4_eneden} shows the result of exact diagonalization for $g = 0.99$ and $g = 1.01$ cases. The vertical axis represents the logarithm of the ground state energy density, and the horizontal axis represents the number of sites $N$. From this figure, we can see that the ground state energy density decreases exponentially as $N$ increases. Hence, we expect that the ground state energy densities for $g = 0.99$ and $g = 1.01$ converge to zero in the thermodynamic limit. Therefore, SUSY is restored in the infinite volume limit when $g = 0.99$ and $g = 1.01$.

\begin{figure}[htb]
\begin{center}
\includegraphics[width=1.0\columnwidth]{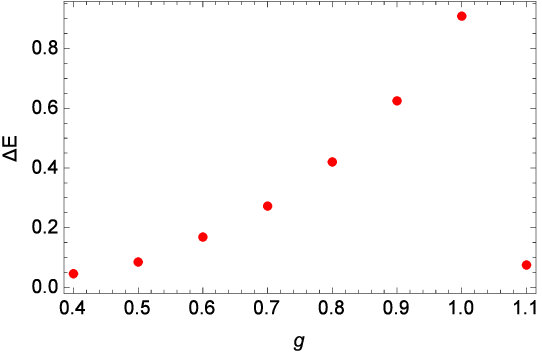}
\end{center}
\caption{The energy gap for $g = 0.4, \dots, 1.1$. The horizontal axis is the parameter $g$.}
\label{fig:chap4_gap}
\end{figure}

In the next section, we prove that there exist gapless modes when SUSY is broken spontaneously by giving an inequality with the same strategy as the previous chapter. By considering the contraposition of the statement, {\ there exists gapless modes when SUSY is spontaneously broken}, reads that {\it SUSY is unbroken in the infinite volume limit when the system is gapped}. To obtain further evidence of restoration of SUSY, we calculated the energy gap for $g = 0.4, \dots, 1.1$. Fig.~\ref{fig:chap4_gap} shows the results of exact diagonalization. The vertical axis represents the energy gap, and the horizontal axis represents the parameter $g$. These values of the energy gap are obtained by finite-size scaling. From the fig.~\ref{fig:chap4_gap}, we find that SUSY is restored in the infinite volume limit. 

\bigskip

\subsection{SUSY breaking\label{sec:chap4_breaking}}
In this subsection, we discuss the condition of spontaneous SUSY breaking. Below we prove that SUSY is spontaneously broken in both finite and the infinite systems. In the case of $g = 1 + \delta$, Eq. (\ref{eq:chap4_Ham}) reads
\begin{align}
H = \frac{N}{2}(\delta^2 + 2\delta) + 2\delta \ii \sum_{j = 1}^N\gamma_j\gamma_{j+1}+H_{g=1},
\label{eq:deltaexp}
\end{align} 
where $H_{g=1}$ is the Hamiltonian with $g=1$.
Using Anderson's argument~\cite{PR_Anderson, PRB_Valenti, PRD_Beccaria, PRL_Nie}, we get a rigorous lower bound of the ground state energy $E_0$ as follows,
\begin{align}
E_0\ge \frac{N}{2}(\delta^2 + 2\delta) + E_0^{\rm free}.
\end{align}
Here, we use the fact that the ground state energy is $0$ when $g = 1$ and $E_0^{\rm free}$ denotes the ground state energy of the second term of right hand side in Eq. (\ref{eq:deltaexp}). The ground state energy of the free part $E_0^{\rm free}$ is calculated as
\begin{align}
E_0^{\rm free}=-\frac{4\delta}{\tan(\pi/N)}.
\label{eq:chap4_freeE}
\end{align}
The derivation of Eq. (\ref{eq:chap4_freeE}) is discussed in appendix~\ref{appB:fourier}. Using the inequality $x<\tan(x)$ for $0<x<\pi/2$, we obtain the following inequality of the ground state energy density,
\begin{align}
\frac{E_0}{N}\ge \frac{\delta}{2}\left(\delta + 2-\frac{8}{\pi}\right).
\label{eq:chap4_LB}
\end{align}
It follows from this inequality that SUSY is spontaneously broken when $g=1 + \delta>8/\pi - 1$ in both finite and the infinite systems. We note that this condition is not optimal. In the appendix~\ref{appB:ged}, numerical calculations show that the ground energy density is positive, in both finite and infinite systems, even when $g$ is smaller than $8/\pi -1$.

\bigskip

\section{Nambu-Goldstone modes}
\label{sec:chap4_NGmodes}
In this subsection, we prove that there exist gapless modes associated with spontaneous SUSY breaking. To prove it, we derive an inequality using a variational method based on Bijl-Feynman ansatz~\cite{PR_Feynman, ZPB_Horsch, PRB_Stringari, JPSJ_Momoi}. Here, we define a variational energy $\epsilon_{\rm var}(p)$ as
\begin{align}
\epsilon_{\rm var}(p) = \frac{\langle\psi_0|Q_pHQ_p|\psi_0\rangle}{\langle\psi_0|Q_p^2|\psi_0\rangle} - E_0.
\end{align}
Here, $|\psi_0\rangle$, $H$, $Q_p$, and $E_0$ are SUSY broken ground state, the Hamiltonian, the Fourier component of the local supercharge, and the ground state energy, respectively. The Fourier component of the local supercharge $Q_p$ is defined as
\begin{align}
Q_p = \sum_{l}\cos(p l)q_l, \quad q_l = g\gamma_{2l} + {\mathrm i}\gamma_{2l -1}\gamma_{2l}\gamma_{2l+1}.
\end{align}
where the momentum $p$ takes values of $4\pi m/N$ ($m\in\mathbb{Z}$). By straightforward calculation, we find that the variational energy is written by a double commutator
\begin{align}
\epsilon_{\rm var}(p) = \frac{\langle[Q_p, [H, Q_p]]\rangle_0}{\langle\{Q_p, Q_p\}\rangle_0}.
\end{align}
Here, $\langle\cdots\rangle_0$ denotes the expectation value in the ground state. By the Pitaevskii-Stringari inequality Eq.~(\ref{eq:PitaevskiiStringari}), we get the following upper bound of variational energy
\begin{align}
\epsilon_{\rm var}(p) \le \sqrt{\frac{f(p)}{g(p)}},
\end{align}
where
\begin{align*}
f(p) & = \langle\{[H,Q_p],[Q_p,H]\}\rangle_0, \\
g(p) & = \langle\{Q_p, Q_p\}\rangle_0.
\end{align*}
Since $q_l$ satisfies the following locality condition
\begin{eqnarray}
\{q_k,q_l\}=\left\{ \begin{array}{ll}
{\rm nonzero} & |l-k| \le 2 \\
0 & ({\rm others}) \\
\end{array} \right.
\end{eqnarray}
$f(p)$, $g(p)$ are even functions of $p$. Since we obtain $f(0) = 0$ and $g(0) = 2E_0$,  the numerator $f(p)$ and the denominator $g(p)$ can be expanded as 
\begin{align*}
f(p) & = CNp^2 + \mathcal{O}(p^4) \quad , \quad (C \neq 0) \\
g(p) & = 2E_0 + \mathcal{O}(p^2),
\end{align*}
when $p$ is small enough. Hence, we get the following upper bound of the variational energy
\begin{align}
\epsilon_{\rm var}(p)\le\sqrt{\frac{C}{2E_0/N}}|p|+\mathcal{O}(p^3).
\label{eq:chap4_NGf}
\end{align}
From Eq. (\ref{eq:chap4_NGf}), we conclude that there exist gapless modes associated with spontaneous SUSY breaking.

\bigskip

\section{Dispersion relation \label{sec:chap4_dispersion} }
\begin{figure}[htb]
\begin{center}
\includegraphics[width=1.0\columnwidth]{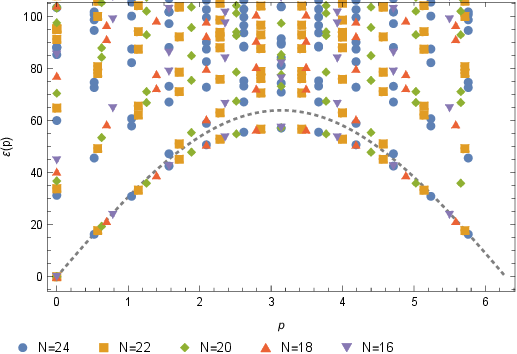}
\end{center}
\caption{Many body spectra for $g = 8$ as a function of momentum $p$. The dotted line denotes the dispersion relation of the free part of the Hamiltonian $H_{\rm free}$.}
\label{fig:chap4_disp}
\end{figure}

In this subsection, we clarify that the dispersion relation of the massless modes proven in the previous subsection have linear dispersion relation in momentum $p$. For the later purpose, let us consider the large $g$ limit of the Hamiltonian. In this limit, $H_{\rm free}$ is dominant to the dispersion relation. By the Fourier transformation, we obtain the following representation of $H_{\rm free}$
\begin{align}
H_{\rm free} = \sum_p \sin(p)\gamma(p)^\dagger\gamma(p).
\end{align}
Here, $\gamma(p)$ is defined as
\begin{align*}
\gamma(p) = \sqrt{\frac{1}{2N}}\sum_{j=1}^N\gamma_j\exp({\mathrm i}pj).
\end{align*}
The Hermitian conjugate of $\gamma(p)$ is defined as $\gamma(p)^\dagger=\gamma(-p)$. From this equation, we find that the dispersion relation in the large-$g$ limit is linear in momentum $p$.
In order to study the dispersion relation of NG fermion, we calculate dispersion relation using exact diagonalization method for $N = 16, 18, 20, 22, 24$. The Figure~\ref{fig:chap4_disp} shows the results of the numerical calculations for $g = 8$. The vertical axis is many-body spectra of the Hamiltonian, and the horizontal axis is momentum $p$. The dotted curve of Fig.~\ref{fig:chap4_disp} corresponds to the one-particle excitation of $H_{\rm free}$. From Fig.~\ref{fig:chap4_disp}, we find that there exist excited states that fit the dotted curve in the vicinity of the zero momentum point. Hence, the lowest-lying states are considered to have gapless and linear dispersion in momentum $p$.

\begin{figure}[htb]
\begin{center}
\includegraphics[width=1.0\columnwidth]{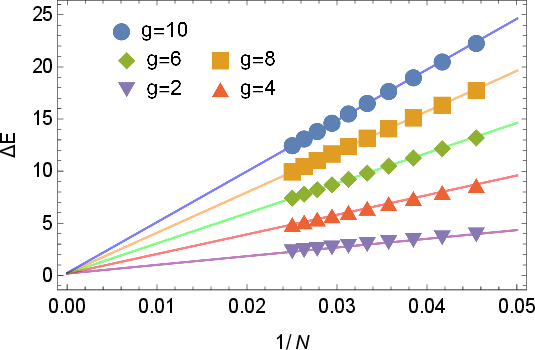}
\end{center}
\caption{Finite size scaling of the first excitation energy of the Hamiltonian for each $g$ as a function of $1/N$. The lines are fits to the data of $N=32, \dots,40$.}
\label{fig:chap4_first}
\end{figure}

As supporting evidence, we also calculate the first excitation energy using exact diagonalization for various values of $g$. Figure~\ref{fig:chap4_first} shows the results of the numerical calculations. The vertical axis is the first excitation energy, and the horizontal axis is $1/N$. The lines are fits to the data of $N = 32, \dots, 40$. From Fig.~\ref{fig:chap4_first}, we find that the first excitation energy is proportional to $1/N$, and converges to $0$ in the large-$N$ limit. Since the momentum $p$ takes the values of $4\pi m/ N$ ($m\in\mathbb{Z}$), we can conclude that the lowest-lying states have gapless and linear dispersion in momentum $p$ when SUSY is spontaneously broken. 

\begin{figure}[htb]
\begin{center}
\includegraphics[width=1.0\columnwidth]{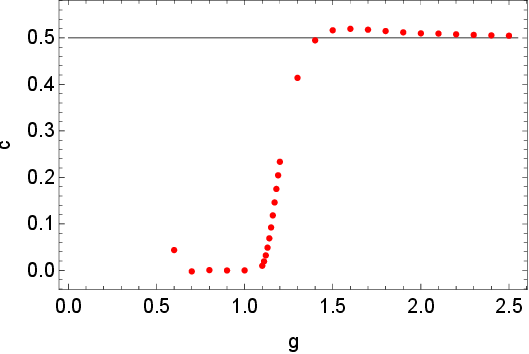}
\end{center}
\caption{The estimated central charge of the model as a function of $g$. The black line represents the central charge of Majorana fermion \makebox{$c = 1/2$}.}\label{fig:chap4_centch}
\end{figure}

By the Jordan-Wigner transformation Eq.~(\ref{eq:JWtr}), the free Hamiltonian $H_{\rm free}$ is mapped to the transverse field Ising model at the critical point
\begin{align*}
H_{\rm free} = \sum_{l}\left(\sigma_l^y\sigma_{l+1}^y - \sigma_{l}^z\right).
\end{align*}
It is well known that critical Ising model is described by CFT with central charge $c = 1/2$. 
We calculate the central charge $c$ in the infinite volume limit by finite size scaling. Figure~\ref{fig:chap4_centch} shows the results of exact diagonalization using the following equation
\begin{align*}
\frac{E_0}{N}=e_{\infty}+\frac{\pi v_Fc}{3N^2}+\mathcal{O}\left(\frac{1}{N^3}\right).
\end{align*}
Here, $E_0$, $e_{\infty}$ and $v_F$ are the ground state energy, the ground state energy density in the large-$N$ limit, and the Fermi velocity, respectively. The Fermi velocity is calculated as $v_{\rm F}=N / 2\pi \Delta E$, and $\Delta E$ is the first excitation energy. The vertical axis of Fig.~\ref{fig:chap4_centch} is the central charge $c$, and the horizontal axis is the parameter $g$. The black line represents $c = 1/2$. From the Fig.~\ref{fig:chap4_centch} , we obtain $c = 1/2$ when $g$ is large enough and $c = 0$ when $g$ is small enough. The central charge $c = 0$ in the case that $g$ is small enough is consistent with the fact that the system is gapped as shown in Sec.~\ref{sec:chap4_restore}. An interacting Majorana model constructed by two supercharges has the central charge $c=7/10$, the same class as the tricritical Ising model, at the critical point of phase transition~\cite{PRL_OBrien}. Nevertheless, this model does not have such an exotic central charge at the critical point. Because of the finite size effect, we cannot determine the critical point of phase transition of SUSY broken phase and SUSY unbroken phase.

\chapter{Summary}
\label{sec:conclusion}
In this chapter, we summarize our results in this thesis. 

In Chap.~\ref{chap:intro}, we first reviewed SSB briefly. Secondly, we reviewed a short history of SUSY. Then, we saw examples of lattice SUSY models with $\mathcal{N}=2$ SUSY in non-relativistic systems focusing on the Nicolai model. We also reviewed NG fermions in the models of the previous study by the author. Finally, we reviewed the ground state properties of the Kitaev chain model.

In Chap.~\ref{chap:N1susy}, we reviewed general properties of $\mathcal{N}=1$ SUSY quantum mechanics, and we gave the definition of spontaneous SUSY breaking.

In Chap.~\ref{chap:MajoNicolai} , we introduced an extension of the Nicolai model using Majorana fermions.  The model interpolates smoothly between the strongly interacting limit and the free Majorana fermion chain as the parameter $|g|$ is varied from 0 to $\infty$. We showed that SUSY is unbroken in both finite and infinite systems for the condition of $|g|=1$. When $|g|$ is small enough, we find that SUSY is restored in the infinite volume limit yet broken in finite systems. For $g>8/\pi - 1$, we proved that SUSY is spontaneously broken in both finite and the infinite systems. We gave an upper bound of the dispersion and showed the existence of gapless modes with a variational argument. This method is general since no symmetry except SUSY is assumed in the proof. We would like to emphasize that the definition of SUSY breaking introduced in Chap~\ref{chap:N1susy} plays an important role in the proof, and, in this sense, the definition is proper. Using numerical exact diagonalization, we found that the gapless excitation has cubic dispersion in momentum.

In Chap.~\ref{chap:LNicolai}, we introduced another model describing interacting Majorana fermions with $\mathcal{N}=1$ SUSY. When $|g|$ is small enough, we find that SUSY is restored in the infinite volume limit yet broken in finite systems. In this case, we also find that the system is gapped. When $g>8/\pi - 1$, we proved that SUSY is spontaneously broken in both finite and the infinite systems. Using numerical exact diagonalization, we found that the gapless excitation has linear dispersion relations and is described by $c=1/2$ conformal field theory.

In non-relativistic supersymmetric quantum mechanics models introduced in this thesis, SUSY is spontaneously broken and SUSY breaking gives rise to gapless excitations, which can be thought of as NG fermions. 

In the author's previous study of lattice fermion models with $\mathcal{N}=2$ SUSY~\cite{U1Nicolai, Z2Nicolai}, the same dispersion relation was obtained when SUSY is spontaneously broken, i.e., linear dispersion and cubic dispersion in momentum. However, in this thesis, we found differences between $\mathcal{N}=2$ SUSY and  $\mathcal{N}=1$ SUSY. In the models with  $\mathcal{N}=2$ SUSY, SUSY is unbroken in one point, i.e. $g=0$, in both finite and the infinite systems. In finite systems, the number of ground states grows exponentially as the system size increases for $g=0$. On the other hand, in $\mathcal{N}=1$ SUSY models, there is an extended area of parameter space that SUSY is restored in the infinite volume limit. In Chap.~\ref{chap:LNicolai}, we numerically showed that there is a gapped area in parameter space when SUSY is unbroken in the infinite volume limit. This result provides that the definition of spontaneous SUSY breaking given in Chap.~\ref{chap:N1susy} is proper since it means that there is no gapless modes without spontaneous SUSY breaking. When SUSY is unbroken in finite systems, i.e. $g=1$, the ground state degeneracy is of order $1$. 

We find that there are differences between relativistic systems and non-relativistic systems. In high-energy physics, superalgebra must hold the Lorentz symmetry. Therefore, the dispersion relation of NG fermions must be linear in momentum. In the model of Chap.~\ref{chap:MajoNicolai}, the dispersion relation of NG fermions is cubic in momentum, which violates the Lorentz symmetry. This cubic dispersion relation implies exotic excitations. 

There are also differences between ordinary spontaneous symmetry breaking and spontaneous SUSY breaking. The counting theories of NG boson, which provide general descriptions of NG boson and broken symmetry, were constructed by Watanabe-Murayama~\cite{PRL_Watanabe}, and Hidaka~\cite{PRL_Hidaka}. If we apply the counting theories of NG bosons to supersymmetry naively, we get quadratic dispersion relations by scaling analysis. In this thesis, we showed that dispersion relations of NG fermions are linear or cubic in momentum. This shows that we cannot apply the counting theory of NG bosons to broken supersymmetry directly and we need to construct a new theory to describe NG fermions in a unified form. As future work, it would be interesting to classify NG fermions in non-relativistic systems and construct a model-independent theory that gives the precise number of NG fermions. In this sense, our studies including $\mathcal{N}=2$ SUSY models provide the first step for a comprehensive understanding and unified descriptions of NG fermions and spontaneous SUSY breaking which are independent of details of systems.


\appendix 



\chapter{Appendix to Chap.~\ref{chap:MajoNicolai}}
\label{chap:appa}


\section{Degenerate ground states\label{appA:dgs}}
In this section, we prove that
\begin{align*}
\sum_{j=1}^Ne^{-{\mathrm i}\frac{\pi}{4}}\gamma_j|\Psi_0\rangle
\end{align*}
is also the ground state of the Hamiltonian Eq.~\ref{eq:chap3_Ham} with $g=1$.
By calculating anti-commutator acting state $|\Psi_0\rangle$, we obtain
\begin{align*}
\{Q,A\}|\Psi_0\rangle=\left(\frac{e^{{\rm i}\pi/4}}{\sin(\pi/8)}+\frac{1+e^{-{\rm i}\pi/4}}{2\sin(\pi/4)}\right)(e^{-{\rm i}\pi L/2}-1)|\Psi_0\rangle
\end{align*}
where the operator A is defined as 
\begin{align*}
A=\sum_{j=1}^Ne^{-{\rm i}\frac{\pi}{4}j}\gamma_j.
\end{align*}
In the case of $L = N/2 = 4m$ ($m$ is an integer), we obtain
\begin{align*}
\{Q,A\}|\Psi_0\rangle=0.
\end{align*}
Since $|\Psi_0\rangle$ is the ground state and annihilated by $Q$, $A|\Psi_0\rangle$ is also annihilated. Threfore $A|\Psi_0\rangle$ is also the ground state of the Hamiltonian for $g = 1$. For the simplicity, we define $|\Phi_0\rangle = A|\Psi_0\rangle$. Since $H$, $T^2$ and $(-1)^F$ are commuting with each other, the ground states are eigenstates of $H$, $(-1)^F$ and $T^2$. Here, $T$ is translational operator of Majorana fermion. By applying $T^2$, we find
\begin{align*}
T^2|\Phi_0\rangle=e^{{\mathrm i}\pi/2}(\cdots+e^{-{\rm i}\pi N/4}e^{-{\rm i}\pi /4}\gamma_1+e^{-{\rm i}\pi N/4}e^{-{\rm i}\pi/2}\gamma_2+\cdots)|\Psi_0\rangle
\end{align*}
Since $|\Phi_0\rangle$ must be the eigenstate of $T^2$, i.e.,$T^2|\Phi_0\rangle\propto|\Phi_0\rangle$, we find
\begin{align*}
e^{-{\rm i}\pi N/4}=1.
\end{align*}
Hence, we get $N = 8m$ ($m \in \mathbb{Z}$) and, in this case , $\sum_{j=1}^Ne^{-{\mathrm i}\frac{\pi}{4}}\gamma_j|\Psi_0\rangle$ is the ground state.

\bigskip

\section{Ground state energy of $H_{\rm free}$\label{appA:free}}
The free part of the Hamiltonian $H_{\rm free}$ can be rewritten as follows
\begin{align*}
H_{\rm free} = \frac{\mathrm i}{4}\Gamma^t {\mathcal H_{\rm free}} \Gamma,
\end{align*}
where $\Gamma$ is an $N$-dimensional vector whose components are $\left(\gamma_1, \gamma_2, \dots , \gamma_N\right)^t$. Here, ${\mathcal H_{\rm free}}$ is an $N\times N$ real skew symmetric matrix whose explicit form is given by
\begin{align}
{\mathcal H_{\rm free}}=g\begin{pmatrix}
~0~ & 4 & -2 & 0 & 0 &  \dots & 0 & 0 & 2 & ~-4~ \\
~-4~ & 0 & 4 & -2 & 0 &  \dots & 0 & 0 & 0 & ~2~ \\
2 & -4 & 0 & 4 & -2 &  \dots & 0 & 0 & 0 & ~0~ \\
\vdots &   &   & \ddots &  &  & \ddots &  &  & ~\vdots~ \\
0 & 0 & 0 & 0 & 0 &  \dots & -4 & 0 & 4 & ~-2~ \\
-2 & 0 & 0 & 0 & 0 &  \dots & 2 & -4 & 0 & ~4~ \\
4 & -2 & 0 & 0 & 0 &  \dots & 0 & 2 & -4 & ~0~ 
\end{pmatrix}.
\end{align}
By employing an orthogonal matrix $Q$, ${\mathcal H_{\rm free}}$ is block diagonalized as
\begin{align}
Q^tAQ = \bigoplus_{j = 1}^{N/2} R_l,
\end{align}
where
\begin{align*}
R_j = \begin{pmatrix}
~0~ & ~\epsilon_j~ \\
~-\epsilon_j~ & ~0~
\end{pmatrix}
\end{align*}
Here, $\pm{\mathrm i}\epsilon_j$ are eigenvalues of $A$. Since eigenvalues of real skew symmetric matrices are pure imaginary, we introduce a new matrix $\tilde{A} := {\mathrm i}A$ to make eigenvalues real. To obtain eigenvalues, we solve the following eigenvalue equation
\begin{align}
\tilde{A} {\bm v} = \lambda {\bm v},
\label{eq:EigenEq}
\end{align}
where ${\bm v}$ is an $N$ dimensional vector and its components are \makebox{$\left(v_1, v_2, \dots, v_N\right)^t$}. We get the following equations about every component from Eq. (\ref{eq:EigenEq})
\begin{align}
2v_{j-2}-4v_{j-1}+4v_{j+1}-2v_{j+2} = \lambda v_{j} , \quad \left(j = 1, \dots, N \quad \quad {\rm mod} \ N\right).
\end{align}
Since the original Hamiltonian is translationally invariant, we suppose that the eigenvectors are 
plane waves,
\begin{align*}
v_j = C \exp({\mathrm i} p j) .
\end{align*}
We get
\begin{align*}
\epsilon & =g2\mathrm{i}(e^{\mathrm{i}p(j-2)}-2e^{\mathrm{i}p(j-1)}+2e^{\mathrm{i}p(j+1)}-e^{\mathrm{i}p(j+2)})e^{-\mathrm{i}pj} \\
& =-8g\sin(p)+4g\sin(2p).
\end{align*}
The momentum $p$ is an element of the set $\mathcal{K}$. Here, $\mathcal{K}$ is defined as
\begin{align*}
\mathcal{K}=\left\{ 0,\pm\frac{2\pi}{N},\pm\frac{4\pi}{N},\dots,\pm\frac{(N-2)\pi}{N},\pi \right\}.
\end{align*}
The ground state energy $E_0^{\rm free}$ of $H_{\rm free}$ is obtained as follows,
\begin{align*}
E_0^{\rm free} & =-\cfrac12 \sum_{l=1}^{N/2}\left(8g\sin\left(\frac{2 \pi l }{N}\right)-4g\sin\left(\frac{4\pi l}{N}\right)\right) \nonumber \\
 & = -\frac{8g}{\tan\left(\pi/N\right)}.
\end{align*}

\bigskip

\section{Fourier Transform of $H_{\rm free}$\label{appA:Fourier}}
In this section, we calculate the Fourier transform of $H_{\rm free}$ which is defined by
\begin{align*}
H_{\rm free}=2g\mathrm{i}\sum_{j=1}^N(2\gamma_j\gamma_{j+1}-\gamma_{j-1}\gamma_{j+1}).
\end{align*}
Here, we assume PBC and 
$N$ is even. The Fourier transform of Majorana fermion operators are calculated as~\cite{Rahmani_PRB}
\begin{align*}
\gamma_j=\sqrt{\frac{2}{N}}\sum_{p}\gamma(p)e^{\mathrm{i}pj},
\end{align*}
and the Inverse Fourier transform is also defined by
\begin{align*}
\gamma(p)=\sqrt{\frac{1}{2N}}\sum_{j=1}^N\gamma_je^{-\mathrm{i}pj}.
\end{align*}
Here, $p$ takes the values of $2\pi m/N$~($m\in \mathbb{Z}$) since PBC is assumed. From the Clifford algebra of Majorana fermion operators, one finds that Fourier transform of Majorana operators satisfy the following anti-commutation relation
\begin{align*}
\{\gamma(p),\gamma(p')\}=\delta_{p,-p'}. \quad , \quad \gamma(-k)=\gamma^\dagger(k).
\end{align*}
By the Fourier transformation, the free part of the Hamiltonian $H_{\rm free}$ reads
\begin{align*}
H_{\rm free} & =2g\mathrm{i}\sum_{j=1}\left(2\gamma_j\gamma_{j+1}-\gamma_{j-1}\gamma_{j+1}\right) \nonumber \\
& =\sum_{p>0}\left(2\sin(p)-\sin(2p)\right)\gamma(-p)\gamma(p)-8g\sum_{p>0}\sin(p).
\end{align*}
By a straightforward calculation, we obtain the following equation
\begin{align*}
H_{\rm free} =8g\sum_{p>0}(2\sin(p)-\sin(2p))\gamma^\dagger(p)\gamma(p)-\frac{8g}{\tan(\pi/N)}.
\end{align*}
The last constant term is the same as the ground state energy as calculated in Sec.~\ref{appA:free}.
From this, we find that the dispersion relation is cubic in momentum when the momentum $p$ is small enough, i.e.,
\begin{align*}
8g(2\sin(p)-\sin(2p))\sim 8g|p|^3.
\end{align*}

\section{Finite size scaling of the ground state energy density for $g\le8/\pi - 1$\label{appA:ged}}
\begin{figure}[htb]
\begin{center}
\includegraphics[width=1.0\columnwidth, clip]{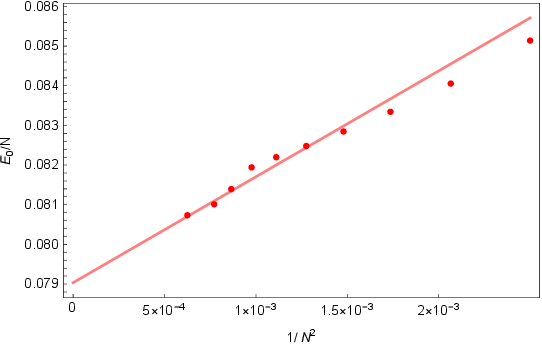}
\end{center}
\caption{Finite size scaling of the ground state energy density for $g = 1.5$ with $N = 22, \dots, 40$. The red line is a fit to the data for $N = 32, \dots, 40$.}
\label{fig:appA_Eden}
\end{figure}

In Sec. \ref{sec:chap3_broken}, we have proved that the ground state energy density is positive both finite and the infinite systems when $g > 8 / \pi - 1$. In this appendix, we would like to present that, even when $g$ is smaller than $8/\pi - 1 \approx 1.546$, the ground state energy density may be positive. Figure \ref{fig:appA_Eden} shows the plot of the data for $g = 1.5$ case calculated by exact diagonalization. The vertical axis represents the ground state energy density, and the horizontal axis represents $1 / N^2$. The red line is a fit to the data for $N = 32, \dots, 40$. Even though $g$ is smaller than $8/\pi-1$, the $N$-dependence of the ground state energy density is the same as free part since the ground state energy density of $H_{\rm free}$ scales as follows in the large-$N$ limit,
\begin{align*}
\frac{E_0^{\rm free}}{N} & = -\left(\frac{1}{N}\right)\frac{4g}{\tan(\pi / N)} = -\frac{4g}{\pi} + \frac{4g\pi}{3N^2} + \mathcal{O}(1 / N^4).
\end{align*}
When $g = 1.5$, the ground state energy density in the infinite volume limit is calculated as $0.079\dots$, so that SUSY is spontaneously broken. Hence, even when $g$ is smaller than $8/\pi-1$, there is a possibility that SUSY is spontaneously broken in finite and the infinite systems.

\section{The dispersion relation is not linear\label{appA:notlinear}}
\begin{figure}[htb]
\begin{center}
\includegraphics[width=1.0\columnwidth]{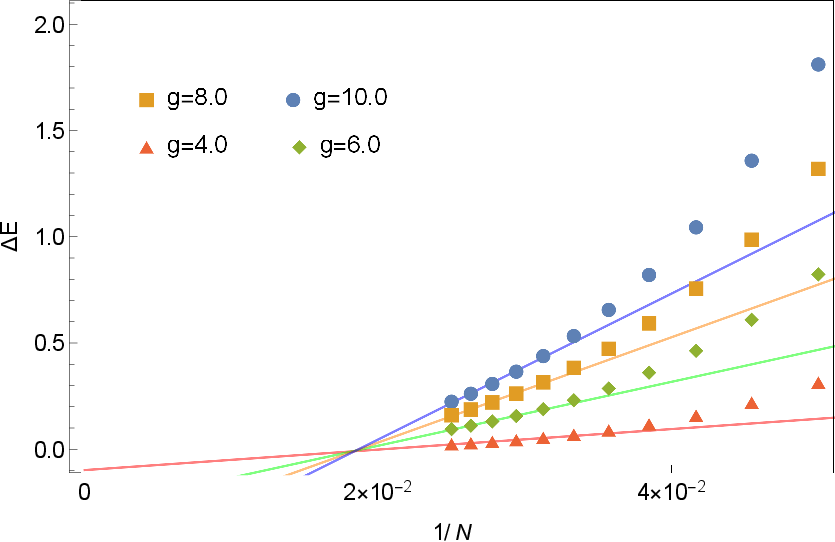}
\end{center}
\caption{Finite size scaling of the first excitation energy of the Hamiltonian for each $g$ as a function of $1/N$. The lines are fits to the data of $N=32, \dots,40$.}
\label{fig:appa_first_linear}
\end{figure}

\begin{figure}[htb]
\begin{center}
\includegraphics[width=1.0\columnwidth]{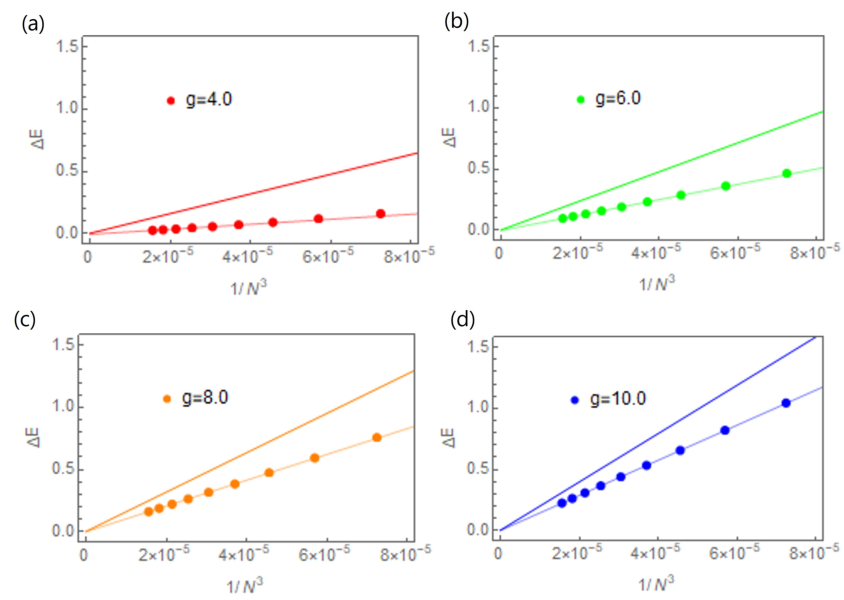}
\end{center}
\caption{(a) Finite size scaling of the first excitation energy for $g=4$ as a function of $1/N^3$. (b) Finite size scaling of the first excitation energy for $g=6$ as a function of $1/N^3$. (c) Finite size scaling of the first excitation energy for $g=8$ as a function of $1/N^3$. (d) Finite size scaling of the first excitation energy for $g=10$ as a function of $1/N^3$. In Figs. (a),(b), (c), (d),  the upper lines stand for $8gp^3$ and the lower lines are fits to the data of $N=32, \dots,40$.}
\label{fig:appa_first_all}
\end{figure}

\begin{figure}[htb]
\begin{center}
\includegraphics[width=1.0\columnwidth]{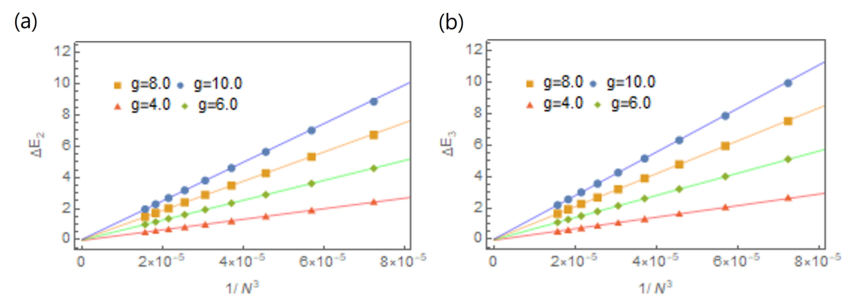}
\end{center} 
\caption{(a) Finite size scaling of the second excitation energy relative to the ground state energy for $g=4, \dots, 10$ as a function of $1/N^3$. (b) Finite size scaling of the third excitation energy relative to the ground state energy for $g=4, \dots, 10$ as a function of $1/N^3$. In both figures, the lines are fits to the data of $N=32, \dots,40$.}
\label{fig:appa_second_third}
\end{figure}

In Sec.~\ref{sec:chap3_dispersion}, we numerically find that the dispersion relation of NG fermions is cubic in momentum. As further support, in this section, we provide other numerical results that show the dispersion relation of NG fermions is cubic in momentum. The figure~\ref{fig:appa_first_linear} shows the result of exact diagonalization, and is the first excitation energy as a function of $1/N$ for $g= 4, 6, 8, 10$. The symbol $\Delta E$ stands for the first excitation energy relative to the ground state energy. The lines are fits to the data of $N=32, \dots, 40$. From this figure, we find that the intercept is minus for $g= 6, 8, 10$, which means that the first excitation energy relative to the ground state energy $\Delta E$ is minus. This is contradictory. Hence, we clarify that the dispersion relation of NG modes is cubic in momentum, i.e., the linear fitting is wrong.
The figure shows the results of exact diagonalization for $g = 4, \dots, 10$. In the graph (a), \dots, (d), the upper lines represents $8gp^3$ which is leading order of the one-particle excitation of $H_free$, i.e., $8gf(p)$. In short, the upper lines correspond to the dotted curve in the vicinity of the point $p=0$. The lower lines are fits to the data of $N=32, \dots, 40$. Since we assume the periodic boundary conditions in the system, the momentum $p$ of the first excitation takes the value of  $p=2\pi/N$. The slope of the upper lines in the Fig.\ref{fig:appa_first_all} is $64\pi^3g\sim1984.4 g$. The slopes of the graph (a), (b), (c), and (d) can be calculated numerically and is $2029.111\dots$, $6314.001\dots$, $10397.580\dots$ and $14408.714\dots$, respectively. Dividing by $g$, we obtain $507.278\dots$, $1052.3\dots$,$1299.7\dots$ and $1440.871\dots$, respectively. From these results, we find that the slope of the lower line divided by $g$ is getting close to that of the upper line ($64\pi^3$) as $g$ increases. This is consistent with the fact that the free part of the Hamiltonian is dominant when $g$ is large enough as seen in the Sec.~(\ref{sec:chap3_dispersion}). Hence, the dispersion of one-particle excitation of $H_{\rm free}$, i.e. $8gf(p)$, gives the upper bound of the cubic dispersion.
The figure~\ref{fig:appa_second_third} provides the results of exact diagonalization for $g=4, 6, 8, 10$. In the graph (a) of Fig.~\ref{fig:appa_second_third}, the vertical axis $\Delta E_2$ represents the second excitation energy relative to the ground state energy, and the horizontal axis is $1/N^3$. In the graph (b) of Fig.~\ref{fig:appa_second_third}, the vertical axis $\Delta E_3$ represents the third excitation energy relative to the ground state energy, and the horizontal axis is $1/N^3$. The lines are fits to the data of  $N=32, \dots,40$. From the figure~\ref{fig:appa_second_third}, we find that the second and the third excitation energies are proportional to $1/N^3$, and these excitations are considered to be many-particle bound states.


\chapter{Appendix to Chap.~\ref{chap:LNicolai}}
\label{chap:appb}


\section{The Ground-sate energy of $H_{\rm free}$}
\label{sec:gsenergy}
The free part of the Hamiltonian $H_{\rm free}$ is given as follows,
\begin{align}
H_{\rm free}=2g\mathrm{i}\sum_{j=1}^{N}\gamma_j\gamma_{j+1}.
\label{eq:ham1}
\end{align}
Here, we assume periodic boundary conditions (PBC). The ground state energy of $H_{\rm free}$ can be calculated exactly since it is a quadratic form. First of all, we rewrite the Hamiltonian as follows,
\begin{align*}
H_{\rm free} & =  2g\mathrm{i} \sum_{j=1}^{N}(\gamma_j\gamma_{j+1}) = \frac{g\mathrm{i}}{2}\Gamma^t A \Gamma,
\end{align*}
Here, $\Gamma$ is a vector whose components are $(\gamma_1, \gamma_2,\dots, \gamma_N)^t$ and $A$ is an $N\times N$ real skew matrix given by
\begin{align*}
A=\begin{pmatrix}
~0~ & 2 & 0 & 0 & 0 &  \dots & 0 & 0 & 0 & ~-2~ \\
~-2~ & 0 & 2 & 0 & 0 &  \dots & 0 & 0 & 0 & ~0~ \\
0 & -2 & 0 & 2 & 0 &  \dots & 0 & 0 & 0 & ~0~ \\
\vdots &   &   & \ddots &  &  & \ddots &  &  & ~\vdots~ \\
0 & 0 & 0 & 0 & 0 &  \dots & -2 & 0 & 2 & ~0~ \\
0 & 0 & 0 & 0 & 0 &  \dots & 0 & -2 & 0 & ~2~ \\
2 & 0 & 0 & 0 & 0 &  \dots & 0 & 0 & -2 & ~0~ 
\end{pmatrix}.
\end{align*}
Since the matrix $A$ is real skew symmetric, it can be transformed into a block diagonal matrix using an orthogonal matrix $O$,
\begin{align*}
O^tAO=\bigoplus_{j=1}^M\begin{pmatrix}
~0~ & \epsilon_j  \\
~-\epsilon_j~ & 0 
\end{pmatrix}.
\end{align*}
Here, $\pm\mathrm{i}\epsilon_j$ are eigenvalues of $A$, and we assume that each $\epsilon_j$ is non-negative. Similary to  the Kitaev chain~\cite{Kitaev_2001}, the ground state energy of the Hamiltonian Eq.~(\ref{eq:ham1}) can be calculated as
\begin{align*}
E^{\rm free}_0=-g\sum_{j=1}^{N/2}\epsilon_j.
\end{align*}
Next, let us consider the eigenvalues of the matrix $A$. For simplicity, we introduce the $N\times N$ Hermitian matrix $\tilde{A}$ as follows,
\begin{align*}
\tilde{A}=\mathrm{i}A.
\end{align*}
We consider the following eigenvalue problem,
\begin{align*}
\tilde{A}\bm{v}=\epsilon\bm{v}.
\end{align*}
Each component $v_j \ (j=1,\dots, N)$ satisfies the following equations,
\begin{align*}
2\mathrm{i}(-v_{N-1}+v_{1}) & =\epsilon v_{N} \nonumber \\
2\mathrm{i}(-v_{N}+v_{2}) & =\epsilon v_{1}\nonumber \\
2\mathrm{i}(-v_{j-1}+v_{j+1}) & =\epsilon v_{j} \quad (j=2,\dots, N-1). 
\end{align*}
Here, we suppose the next ansatz,
\begin{align*}
v_j=\alpha e^{{\mathrm i}pj}.
\end{align*}
From the result of $j=2,\dots, N-1$ case, we get
\begin{align*}
\epsilon & =2\mathrm{i}(e^{\mathrm{i}p(j+1)}-e^{\mathrm{i}p(j-1)})e^{-\mathrm{i}pj} \\
& = 2\mathrm{i}(e^{\mathrm{i}p}-e^{-\mathrm{i}p})=-4\sin(p).
\end{align*}
From the result of $j=1$ case, we get
\begin{align*}
\epsilon=2\mathrm{i}(e^{\mathrm{i}p(2)}-e^{\mathrm{i}p(N)})e^{-\mathrm{i}p}=2\mathrm{i}(e^{\mathrm{i}p}-e^{-\mathrm{i}p}\cdot e^{\mathrm{i}pN}).
\end{align*}
Since $\tilde{A}$ is a Hermitian matrix, the eigenvalues $\epsilon$ must be real ($\epsilon^\ast=\epsilon$). This condition leads to the periodic boundary conditions $e^{\mathrm{i}pN}=1$. Using this condition, we get $\epsilon=-4\sin(p)$ ($p\in\mathcal{K}$) for all $j=1,\dots,N$. Here, $\mathcal{K}$ is the following set
\begin{align*}
\mathcal{K}=\left\{ 0,\pm\frac{2\pi}{N},\pm\frac{4\pi}{N},\dots,\pm\frac{(N-2)\pi}{N},\pi \right\}.
\end{align*}
The ground state energy $E_0^{\rm free}$ of $H_{\rm free}$ is obtained as follows,
\begin{align*}
E_0^{\rm free} & =-g\sum_{n =1}^{N/2}4\sin\left(\frac{2\pi}{N}n\right) =-\frac{4g}{\tan(\pi/N)}.
\end{align*}

\section{Frourier transformation of $H_{\rm free}$\label{appB:fourier}}
In this appendix, we carry out the Fourier transformation of the free part of the Hamiltonian $H_{\rm free}$. The Fourier transformation of the Majorana fermion operator $\gamma_j$ is defined by
\begin{align*}
\gamma_j=\sqrt{\frac{2}{N}}\sum_{-\pi< k\le\pi}e^{-\mathrm{i}kj}\gamma(k).
\end{align*}
Here, $k$ is a momentum and takes the values of $2\pi m/N$ with PBC. The inverse Fourier transformation is defined as follows,
\begin{align*}
\gamma(k)=\sqrt{\frac{1}{2N}}\sum_{j=1}^Ne^{\mathrm{i}kj}\gamma_j.
\end{align*}
This transformation indicates the following relations,
\begin{align*}
\{\gamma(k),\gamma(k')\}=\delta_{k,-k'} \quad , \quad \gamma(-k)=\gamma^\dagger(k).
\end{align*}
With the Fourier transformation, $H_{\rm free}$ is transformed as follows
\begin{align*}
H_{\rm free} & = 2g\mathrm{i}\sum_{j=1}^N\gamma_j\gamma_{j+1} \\
& =4g\mathrm{i}\sum_{k}e^{-\mathrm{i}k}\gamma(-k)\gamma(k).
\end{align*}
Here, we use the following relation, $\frac{1}{N}\sum_{j=1}^Ne^{-\mathrm{i}(k'+k)j}=\delta_{k,-k'}$. By straight forward calculations, we obtain
\begin{align*}
H_{\rm free}=4g\sum_k\sin(k)\gamma(-k)\gamma(k).
\end{align*}
By taking the momentum $k$ positive, we get the following representation of the Hamiltonian,
\begin{align*}
H_{\rm free}=8g\sum_{0\le k\le\pi}\sin(k)\gamma^\dagger(k)\gamma(k)-4g\sum_{0\le k\le\pi}\sin(k).
\end{align*}
By straightforward calculation, we obtain the following representation of the Hamiltonian
\begin{align*}
H_{\rm free}=8g\sum_{0\le k\le\pi}\sin(k)\gamma^\dagger(k)\gamma(k)-\frac{4g}{\tan(\pi/N)}.
\end{align*}
The last constant term coincides with the ground state energy of the Hamiltonian. From this, we can see that the dispersion relation is linear in momentum when the momentum $p$ is small enough.

\section{Finite size scaling of the ground state energy density for $g\le8/\pi-1$\label{appB:ged}}
\begin{figure}[htb]
\begin{center}
\includegraphics[width=0.9\columnwidth, clip]{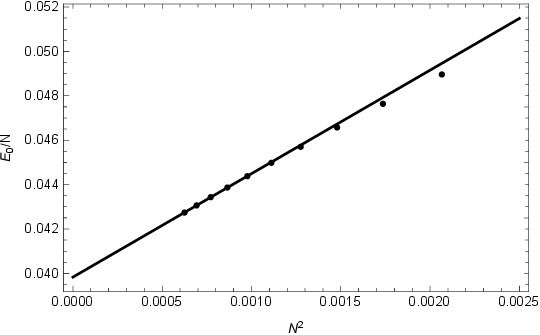}
\end{center}
\caption{Finite size scaling of the ground state energy density for $g = 1.5$ with $N = 22, \dots, 40$. The black line is a fit to the data for $N = 32, \dots, 40$.}
\label{fig:appB_Eden2}
\end{figure}

In Sec. \ref{sec:chap4_breaking}, we have proved that the ground state energy density is positive both finite and the infinite systems when $g > 8 / \pi - 1$. In this appendix, we would like to present that, even when $g$ is smaller than $8/\pi - 1 \approx 1.546$, the ground state energy density may be positive. Figure \ref{fig:appB_Eden2} shows the plot of the data for $g = 1.5$ case calculated by exact diagonalization. The vertical axis represents the ground state energy density, and the horizontal axis represents $1 / N^2$. The black line is a fit to the data for $N = 32, \dots, 40$. Even though $g$ is smaller than $8/\pi-1$, the $N$-dependence of the ground state energy density is the same as free part since the ground state energy density of $H_{\rm free}$ scales as follows in the large-$N$ limit,
\begin{align*}
\frac{E_0^{\rm free}}{N} & = -\left(\frac{1}{N}\right)\frac{4g}{\tan(\pi / N)} = -\frac{4g}{\pi} + \frac{4g\pi}{3N^2} + \mathcal{O}(1 / N^4).
\end{align*}
For $g = 1.5$, the ground state energy density in the infinite volume limit is calculated as $0.03983\dots$, so that SUSY is spontaneously broken. Hence, even when $g$ is smaller than $8/\pi-1$, there is a possibility that SUSY is spontaneously broken in finite and the infinite systems.



\printbibliography[heading=bibintoc]


\end{document}